\documentclass[a4paper,UKenglish,cleveref, autoref, thm-restate]{lipics-v2021}

\usepackage{graphicx} 
\usepackage{amsfonts,amsmath,amssymb,amsthm}
\newtheorem{assumption}{Assumption}
\usepackage{float}
\usepackage{comment}
\usepackage{mathtools} 

\DeclareMathOperator{\Rem}{Rem}

\DeclareMathOperator{\Gap}{Gap}

\hideLIPIcs  

\title{Splitting for Stability: a Modified Reward Scheme for Transaction-Fee-Only Regimes} 

\author{Eric Brigham}{University of Minnesota, United States} {brigh169@umn.edu}{}{}

\author{Baigang Chen}{University of Minnesota, United States} {chen9464@umn.edu}{}{}

\author{Nicholas Hopper}{University of Minnesota, United States} {hoppernj@umn.edu}{}{}

\authorrunning{Brigham \textit{et al.}} 

\Copyright{Jane Open Access and Joan R. Public} 



\category{} 

\relatedversion{} 

\nolinenumbers 

\EventEditors{John Q. Open and Joan R. Access}
\EventNoEds{2}
\EventLongTitle{42nd Conference on Very Important Topics (CVIT 2016)}
\EventShortTitle{CVIT 2016}
\EventAcronym{CVIT}
\EventYear{2016}
\EventDate{December 24--27, 2016}
\EventLocation{Little Whinging, United Kingdom}
\EventLogo{}
\SeriesVolume{42}
\ArticleNo{23}

\begin{document}

\maketitle

\begin{abstract}
As Bitcoin's block subsidy declines over time, miner revenue will come to depend increasingly on transaction fees. In fee-dominated regimes, winner-take-all reward structures can create incentives for undercutting and fork-based mining strategies, leading to instability and reduced transaction throughput. 

To prevent this instability, we propose a reward-splitting mechanism in which the winning block producer receives an $\alpha$-portion of the block reward, while the rest is distributed among other miners; this incentivizes mining on previous blocks with higher rewards, while preserving the incentive to produce blocks with high rewards.

We evaluate the proposed reward structure through analysis and simulation. Our analysis derives a local profitability condition for one-step Function-Fork deviations, showing how the winner-side reward, loser-side subsidy, and honest hash power jointly determine whether undercutting is profitable. Simulations show parameter regions in which honest behavior dominates fork-based deviations. In the cases where forking becomes dominant, the learned strategy is the least aggressive variant studied. The analysis and simulations suggest that suitable reward-splitting parameters can reduce the honest participation needed to suppress aggressive forking strategies relative to the winner-take-all structure studied in prior work. Finally, we show that the modified reward scheme does not appear to worsen selfish mining incentives.

Together, these results suggest that reward redistribution can mitigate fee-driven fork incentives without introducing stronger selfish mining incentives, providing a possible direction for incentive design in fee-era proof-of-work systems.As Bitcoin's block subsidy declines over time, miner revenue will come to depend increasingly on transaction fees. In fee-dominated regimes, winner-take-all reward structures can create incentives for undercutting and fork-based mining strategies, leading to instability and reduced transaction throughput. 

To prevent this instability, we propose a reward-splitting mechanism in which the winning block producer receives an $\alpha$-portion of the block reward, while the rest is distributed among other miners; this incentivizes mining on previous blocks with higher rewards, while preserving the incentive to produce blocks with high rewards.

We evaluate the proposed reward structure through analysis and simulation. Our analysis derives a local profitability condition for one-step Function-Fork deviations, showing how the winner-side reward, loser-side subsidy, and honest hash power jointly determine whether undercutting is profitable. Simulations show parameter regions in which honest behavior dominates fork-based deviations. In the cases where forking becomes dominant, the learned strategy is the least aggressive variant studied. The analysis and simulations suggest that suitable reward-splitting parameters can reduce the honest participation needed to suppress aggressive forking strategies relative to the winner-take-all structure studied in prior work. Finally, we show that the modified reward scheme does not appear to worsen selfish mining incentives.

Together, these results suggest that reward redistribution can mitigate fee-driven fork incentives without introducing stronger selfish mining incentives, providing a possible direction for incentive design in fee-era proof-of-work systems.

\end{abstract}

\section{Introduction}
\label{sec-introduction}
Bitcoin and similar proof-of-work (PoW) systems are maintained, extended, and propagated by network entities known as \textit{miners}. Indeed, the security of these systems rests largely on the behavior of the miners. In classical PoW systems, miners are incentivized to maintain the system through the financial rewards attached to block production. As mining requires significant financial investment in specialized hardware, miners are typically considered to be rational agents seeking to maximize revenue. Consequently, any deviant strategy which allows miners to outperform the honest/default mining protocol presents a serious threat to the security of a PoW system.

Block production presents miners with an opportunity to capture two types of financial rewards: newly minted currency and transaction fees. As deflationary blockchain systems progress, there are ``halving'' events wherein the amount of currency minted for each successfully mined block is cut in half at a prescribed rate \cite{nakamoto2008bitcoin}. When the amount of newly minted currency reaches zero, the sole (financial) incentive to mine blocks comes from transaction fees. While it was posited by Nakamoto that this would be sufficient, it was discovered in later work that this reward structure can warp the behavior of rational miners \cite{carlsten2016instability}. Specifically, in the transaction fee only reward structure, a new ``Petty Compliance'' strategy comes to dominate honest mining. Put simply, the Petty Compliant miner prefers to mine on blocks which leave larger transaction fees available to them. In the case of a fork, they will mine on the block with \textbf{lower} transaction fees, as this leaves more transaction fees available to them to include in their own block. Once it is observed that some fraction of miners are Petty Compliant, other miners may begin intentionally forking existing blocks with blocks containing \textit{lower} transaction fees. Such behavior is referred to as ``undercutting''. Once a fork exists, rational miners will prefer to mine on the head of blocks produced by undercutting, as they may capture the remaining fees. This leads to problems such as increased fork rate and lower throughput. In extreme cases, the system may become functionally unusable as transaction throughput reaches a standstill. 



The key observation motivating our work is that Petty Compliance and Undercutting are profitable largely because the miner of a block can claim all transaction fees included in that block. We therefore ask whether modifying the reward structure can reduce the profitability of these deviant strategies. In this work, we propose a new reward mechanism for transaction-fee-only regimes. Instead of awarding the entire block reward to the successful miner, our scheme splits each block reward between the winning miner and a set of participating \textit{atomic} miners. The goal of this scheme is to reduce the appeal of undercutting by rewarding the miners for mining on the head of blocks with large transaction fees, while also rewarding miners for including as many fees (and hence transactions) as possible in their blocks.

We support the effectivness of this proposed reward mechanism through both analysis and simulation. Our results suggest that the modified reward structure can produce settings in which honest mining dominates the deviant strategies identified by Carlsten {\em et al.} \cite{carlsten2016instability}. More broadly, our work explores whether reward redistribution can be used as a mechanism for improving incentive compatibility in transaction-fee-only proof-of-work systems.

\paragraph*{Contribution 1: Novel Reward Structure}
We propose a reward structure that splits each block reward between the winning miner and the set of all other \textit{atomic} miners. This structure is designed to disincentivize forking by reducing the immediate reward earned by the winning miner, while subsidizing atomic miners who participate in the protocol.

\paragraph*{Contribution 2: Analysis}
We analyze the incentive effects of the proposed reward structure by comparing honest publication against a one-step Function-Fork deviation. Our analysis first recovers the canonical \(2/3\) undercutting threshold of Carlsten {\em et al.}~\cite{carlsten2016instability} under their stylized exponential model, then extends this style of calculation to our atomic-miner reward split. This yields private profitability thresholds that clarify how the winner-side reward, loser-side subsidy, and optional smoothing component affect the incentive to undercut.

\paragraph*{Contribution 3: Simulation}
We implement the proposed reward structure in a mining simulator based on the framework of Carlsten {\em et al.}~\cite{carlsten2016instability}. We evaluate the policy across multiple reward-splitting rules, including even and proportional redistribution.  We additionally implement a normalization or smoothing of rewards across blocks as proposed by Budinsk{\`y} {\em et al.}~\cite{budinsky2023fee}. Our simulations measure the resulting profitability thresholds and show that suitable reward-splitting parameters can shift incentives toward honest mining.

\section{Background and Related Works}
\label{sec:BackgroundRL}

\subsection{Proof-of-Work Mining, Rewards, and Forking}
\label{subsec:pow-fees}

A blockchain can be conceptualized as a sequence of blocks, with each block referencing the one preceding it. 
Ideally, miners work to extend the current head of the chain; in a ``proof-of-work'' (PoW) blockchain, miners compete to produce valid blocks by expending computational resources. 
For example, in Bitcoin \cite{nakamoto2008bitcoin}, each miner work to solve a cryptopuzzle based on the previous block, and publish the solution before other miners.  
Since the expected time to solve this puzzle is determined by the computational power (hash power) of the miner, a miner's probability to win a block is proportional to its share of the total computational power on the network.  
Solving the puzzle entitles the winning miner to a ``block reward'' consisting of newly minted currency as well as fees paid to the block miner by the transactions included in the block. 
As a deflationary currency, the amount of newly minted Bitcoin with each block will eventually reach zero, leaving only transaction fees for the reward. 
This is sometimes referred to as ``fee-only regimes''. 

During the mining competition, it is possible for two (or more) miners to produce different valid blocks extending the same parent block. 
This event is referred to as a \textbf{fork}: two competing branches of the blockchain with the same history up to the parent block, but \textit{different} blocks at the head.  
A fork presents a miner with a choice of which block to extend. 
In Bitcoin style PoW, this is decided by the \textbf{longest chain rule}: honest miners work to extend the longest valid chain. 
In the case of a tie, Bitcoin dictates that miners should mine on the head of the block that they learned of \textit{first}. 

As miners expend resources in order to compete in the mining process, they are usually considered to be rational agents seeking to maximize their expected revenue. A miner that follows the prescribed mining protocol is referred to as \textit{honest} or \textit{Default Compliant}. Any miner pursuing a strategy which departs from this we refer to as \textit{deviant}. Under typical Bitcoin operation, when newly minted coins are still being added, there are numerous deviant mining strategies which have been studied, e.g. selfish mining \cite{eyal2014majority}. 

Carlsten {\em et al.} \cite{carlsten2016instability} demonstrated that fee-only regimes admit new profitable fork-based deviant strategies, because transaction fees are not independent of miner behavior. 
A miner who includes fewer transactions in their block leaves more transactions available for future blocks, influencing which branch of a fork rational miners will prefer to extend.
This observation leads to the \textit{Petty Compliance} strategy. A Petty Compliant miner adheres to the longest-chain rule, but in the case of a fork prefers the branch that leaves more transaction fees available. In other words, if one branch contains a block with \textit{lower} transaction fees, then a Petty Compliant miner prefers to extend that branch as it gives access to higher remaining transaction fees for its own block. 

The profitability of Petty Compliance creates an opportunity for another deviation: \textit{Undercutting}. An Undercutting miner intentionally creates a fork by publishing a competing block that claims fewer transaction fees than the block it seeks to replace. By leaving more fees unclaimed on its branch, the undercutter makes that branch more attractive to Petty Compliant miners. If the undercutter's block is extended first and wins the fork race, then the Longest-Chain Rule causes honest miners to adopt the undercut branch, allowing the undercutter's block to become part of the canonical chain.

Carlsten {\em et al.} model undercutting behavior in their simulator using a family of strategies called \textit{Function-Fork}. A Function-Fork miner follows a fee-claiming rule that determines what fraction of the available transaction fees to claim when creating a fork.  More aggressive Function-Fork strategies leave more fees behind, increasing the chance to be extended but decreasing the immediate rewards if the fork succeeds.

These deviations are harmful because they increase fork rate and reduce transaction throughput. Instead of miners consistently extending the current chain tip and including available transactions, miners strategically create and/or extend lower-fee forks. The reward policy proposed in this work is designed to reduce the profitability of these deviant strategies by weakening the winner-take-all nature of transaction fee rewards.

\subsection{Reward Redistribution Mechanisms}
\label{subsec:reward-redistribution}
The instability identified by Carlsten {\em et al.} \cite{carlsten2016instability} arises in part because the transaction-fee-only regime is winner-take-all at the block level: the miner of a canonical block receives the transaction fees included in that block, while other miners receive nothing. This creates a direct incentive for miners to manipulate \textit{which} fees are included and \textit{which} branch future miners prefer to extend. A natural response is to ask whether modifying the reward-allocation rule can reduce the profitability of such deviations. 

Several lines of prior work study alternatives to simple winner-take-all reward allocation. Chen {\em et al.} \cite{chen2019axiomatic} study block reward allocation from an axiomatic perspective. Their results provide a formal foundation for understanding when proportional allocation is justified. Other work considers reward sharing in pool-based settings. Kiayias {\em et al.} \cite{kiayias2025pool} model mining pools as oceanic games with both atomic and non-atomic players, comparing proportional sharing with Shapley-value-based alternatives. Although their focus is reward allocation within pools rather than reward redistribution across miners, their distinction between atomic and non-atomic participants is useful. 

Reward redistribution has also been studied in consensus-level mechanisms. Brunjes {\em et al.} \cite{brunjes2020reward} propose reward sharing schemes for stake pools, including a cap-and-margin mechanism to align incentives and encourage a target number of pools. While this work is in a proof-of-stake setting, it illustrates how reward design can shape participant behavior at the protocol level.


Aumayr \textit{et al.}~\cite{aumayr2026optimal} proposes FairChain and proportional reward splitting (PRS), using lower-difficulty workshares to allocate rewards in proportion to eligible proof-of-work contributions. They establish an approximate equilibrium result against mining coalitions for sufficiently large protocol parameters and evaluate resistance to selfish mining and related attacks using a Markov decision process. Although their theoretical analysis includes transaction fees in a fixed aggregate reward budget, their practical evaluation uses fixed rewards per height and does not model strategic transaction selection or fee-driven undercutting. Our work studies these incentives explicitly in transaction-fee-only regimes, where miners can leave fees unclaimed to attract other miners to a competing branch. The allocation rule also differs: our base scheme reserves a tunable fraction of each canonical block's fees for its producer and redistributes the remainder among eligible non-winning miners. We analyze how this allocation affects undercutting profitability and evaluate the resulting strategy choices of heterogeneous miners, including the effects of smoothing. 

The work most closely related to ours is the fee-redistribution smart contract framework of Budinsk\'y {\em et al.} \cite{budinsky2023fee}. They consider longest-chain PoW systems in the fee-only regime and propose that only a portion of each block's transaction fees be paid immediately to the block producer. The remaining fees are deposited into one or more fee-redistribution smart contracts (FRSCs), which distribute rewards over future blocks. This smooths miner revenue over time and reduces the immediate advantage of manipulating the transaction-fee content of a block. In their simulations, this mechanism substantially lowers (cf. Carlsten {\em et al.} \cite{carlsten2016instability}) the fraction of Default-Compliant miners needed to make undercutting unprofitable. 

Other work also studies protocol modification for transaction-fee-dominated regimes. Siddiqui {\em et al.} ~\cite{siddiqui2020bitcoinf} propose BitcoinF, a protocol aimed at improving fairness and system health as newly minted currency becomes insignificant and miner revenue depends increasingly on transaction fees. Their work is motivated by similar concerns about fee-driven instability, though it focuses on fairness and transaction processing rather than redistributing rewards. Related questions of fee allocation also arise outside longest-chain blockchains. For example, M{\"u}ller {\em et al.} ~\cite{muller2025game} study how different fee allocation rules affect transaction-selection incentives in DAG-based ledgers. Their setting differs from ours: they focus on duplicate transaction inclusion across parallel blocks, whereas we study longest-chain PoW systems and fork based deviations in the fee-only setting. 

Our proposed mechanism is similar in motivation but differs in its allocation target and timing. Rather than redistributing fees only across future block winners, we split each block's reward between the winning miner and a set of eligible non-winning atomic miners. Thus, our mechanism redistributes rewards across participants in the current mining ecosystem, rather than only smoothing rewards over time. We also consider a smoothing extension inspired by Budinsk\'y {\em et al.}~\cite{budinsky2023fee}, in which a portion of each block reward is deferred and redistributed across future block winners.

\section{Model and Proposed Reward Structure}
\label{sec:modelPRS}
We now formalize the reward structure studied in this work. In standard PoW blockchains, the miner of a canonical block receives all transaction fees included in that block. Our proposed mechanism modifies this rule by splitting the block reward between the winning miner and a set of eligible non-winning miners. The intuition behind this is twofold. First, reducing the immediate reward paid to a winning miner weakens the direct benefit of strategically winning a fork race. Consequently, rational miners may be less likely to adopt undercutting strategies when the immediate return from a successful undercut is reduced. Second, proportionally redistributing part of the reward to other eligible miners incentivizes them to mine on blocks with higher rewards, while the proportionally retained reward incentivizes miners to construct blocks with higher rewards as well.

We let $R$ denote the reward associated with a canonical block. In the transaction-fee-only regime considered in this work, $R$ consists entirely and exclusively of transaction fees. Given parameters $\alpha, \beta \in [0,1]$, with $\alpha + \beta = 1$, the winning miner receives the winner-side payment $\alpha R$. The remaining amount $\beta R$ is then redistributed among eligible non-winning atomic miners. We refer to this mechanism as \textit{splitting}. 

\subsection{Atomic Miners}
\label{subsec:atomic-miners}
The redistribution component is paid only to miners designated as \textit{atomic}. Intuitively, atomic miners are miners with sufficient hash power to represent meaningful participation in the mining environment. Restricting redistribution subsidies to atomic miners is intended to avoid allocating rewards to arbitrarily small miners, thereby reducing concerns of free-riding and improving Sybil resistance.

Determining atomicity in a real PoW system is a non-trivial task. In our analysis, we treat the relevant participants as atomic in order to isolate the incentive effects of the reward split. For experiments based on empirical mining distributions, we use a simple threshold: a miner is designated as atomic if its share of global hash power is at least $1\%$. For controlled threshold experiments, atomicity is assigned by the experimental setup so that we may study the effect of reward-splitting parameters without also varying which miners qualify as atomic.

\subsection{Even and Proportional Redistribution}
\label{subsec:even-prop-split}

Let $\mathcal{A}$ denote the set of atomic miners and let $w \in \mathcal{A}$ denote the winning miner of a canonical block. The winner receives $\alpha R$, while the subsidy $\beta R$ is distributed among miners in $\mathcal{A} \setminus \{w\}$. 

We consider two simple rules for distributing the subsidy among non-winning atomic miners: even and proportional.

Under \textit{even splitting}, the subsidy is divided equally among all non-winning atomic miners. Thus, each miner $i \in \mathcal{A} \setminus \{w\}$ receives $\frac{\beta R}{|\mathcal{A}| - 1}$. 

Under \textit{proportional splitting}, the subsidy is divided in proportion to the hash power among the non-winning atomic miners. Let $h_i$ denote miner $i$'s hash power and let $H_{\mathcal{A} \setminus \{w\}} = \sum_{j \in \mathcal{A} \setminus \{w\}} h_j$. Then each miner $i \in \mathcal{A} \setminus \{w\}$ receives $\beta R \cdot \frac{h_i}{H_{\mathcal{A} \setminus \{w\}}}$.  

These two rules are intentionally simple. Although prior work studies more sophisticated reward allocations \cite{kiayias2025pool}, our goal is to isolate the incentive effects of redistributing part of each block reward. Even and proportional splitting provide two natural baselines to study this effect; more sophisticated methods may be useful in future work.

\subsection{Smoothing Extension}
\label{subsec:smoothing-extension}
We also study a smoothing extension inspired by Budinsk\'y {\em et al.} \cite{budinsky2023fee}. This extension augments the splitting mechanism described above by deferring a portion of each block reward and adding it to the reward available in future rounds, beginning at the succeeding block. The resulting reward in each future block is then distributed according to the same splitting rule: an $\alpha$-portion is paid to the winner while the $\beta$-portion is redistributed among eligible non-winning atomic miners. 

Formally, given parameters $\alpha, \beta, \delta \in [0,1]$ with $\alpha + \beta + \delta = 1$, the winner of a block receives the immediate winner-side payment $\alpha R$, eligible non-winning atomic miners share the immediate subsidy $\beta R$, and the remaining $\delta R$ portion is deferred. In our scheme, the deferred amount is distributed evenly across the next $L$ canonical blocks, where $L$ is the "smoothing window". Deferred rewards are added to the rewards available in future rounds and are then distributed using the same winner-side and redistribution rules. 

Importantly, the deferred component is treated as a protocol-determined reward. Once a portion of a block reward is deferred, future miners cannot selectively exclude, undercut, or otherwise manipulate that deferred amount in the way that they may with transaction fees.

Implementing such a smoothing rule in a real PoW system is non-trivial, since it requires the protocol to track deferred rewards and enforce their distribution across future canonical blocks. We do not address these implementation details here. Our use of smoothing is intended to model the incentive effect of reward normalization when combined with splitting. For a more detailed discussion of fee-redistribution contracts and their role in smoothing transaction-fee rewards, we refer the reader to Budinsk\'y {\em et al.} \cite{budinsky2023fee}.

This smoothing component is not intended as a novel mechanism by itself. Rather, we include it to study how splitting interacts with reward normalization across blocks. In the analysis and simulation below, we compare splitting with and without smoothing.

\section{Policy Analysis}
\label{sec:policy-analysis}

This section analyzes the reward policies in Section~\ref{sec:modelPRS}. Throughout, a reward scheme is described by parameters $(\alpha,\beta,\delta)$ with $\alpha,\beta,\delta\ge 0$ and $\alpha+\beta+\delta=1$: the miner of a canonical block receives an $\alpha$-share of its reward, eligible non-winning atomic miners share a $\beta$-portion, and the remaining $\delta$-portion is deferred and released over the next $L$ canonical blocks. This family contains every reward structure discussed in this paper:
\begin{itemize}
\item \emph{winner-take-all} ($\alpha=1$): the transaction-fee-only baseline of Carlsten {\em et al.}~\cite{carlsten2016instability};
\item \emph{deferral-only} ($\beta=0$, $\delta\in(0,1)$): an idealization of the fee-redistribution smart contracts (FRSCs) of Budinsk\'y {\em et al.}~\cite{budinsky2023fee};
\item \emph{splitting-only} ($\delta=0$): the mechanism of Section~\ref{subsec:even-prop-split};
\item \emph{splitting with smoothing} ($\alpha,\beta,\delta>0$): the full mechanism of Section~\ref{subsec:smoothing-extension}.
\end{itemize}

We formulate a single one-step deviation model over an arbitrary fee distribution and prove one profitability theorem for the entire family. The classical $2/3$ threshold is recovered as the exponential instance of Corollary~\ref{cor:wta}, which also yields closed-form thresholds for general fee distributions. Corollary~\ref{cor:deferral} gives an analytical counterpart of the FRSC mechanism. Propositions~\ref{prop:no-local-gain} and~\ref{prop:subsidy-dominated} isolate the local effect of the loser-side subsidy, and Theorem~\ref{thm:smoothing} gives exact comparative statics for the smoothing component.

We emphasize the scope of the analysis at the outset. We compare baseline non-forking play against a one-step Function-Fork deviation at a fixed local chain state, conditional on the focal miner being the next miner whose local action is evaluated. The resulting thresholds are local profitability thresholds for undercutting; they are not statements about total system welfare, long-run stationary revenue, or the full downstream effects of fork instability. Those system-level quantities are measured by simulation in Section~\ref{sec:simulation}, and Section~\ref{subsec:analysis-interpretation} discusses how the two levels of analysis relate.

\subsection{The Local Deviation Model}
\label{subsec:ccs16-toy}

We consider a fixed focal atomic miner $i$ at an opportunity: a local pre-mining decision state in which miner $i$ observes the current chain head and the relevant fee quantities and chooses which block-construction rule to use. Miner $i$ compares two actions. Under baseline non-forking play, miner $i$ mines on the current head under the Longest-Chain Rule~\cite{garay2015bitcoin}, claims the available fees, and publishes immediately. Under one-step Function-Fork (FF) play, in the sense of Carlsten {\em et al.}~\cite{carlsten2016instability}, miner $i$ instead applies the Function-Fork rule: when the local fee state favors undercutting, it mines a competing block on the parent of the current head, deliberately leaving fee mass unclaimed so that the next miner may prefer to extend the fork; otherwise it mines as an ordinary extension.

\begin{definition}[Fee primitives]
\label{def:fee-primitives}
At an opportunity, let $\Rem\ge 0$ denote the fee mass available to an ordinary extension of the current head, and let $\Gap\ge 0$ denote the fee mass that a one-step undercutting block claims while leaving the remainder unclaimed to attract the next miner. The pair $(\Rem,\Gap)$ has an arbitrary joint distribution with finite means and $\Pr[\Rem=\Gap]=0$. Define \(\eta_R:=\mathbb E\bigl[\Rem\,\mathbf 1\{\Rem>\Gap\}\bigr]\), \(\eta_G:=\mathbb E\bigl[\Gap\,\mathbf 1\{\Gap>\Rem\}\bigr]\), \(p_G:=\Pr[\Gap>\Rem]\), together with the baseline fee mass $\mu_i^{\mathrm{opp}}:=\mathbb E[\Rem]$ and the undercutting surplus $\Delta:=\mathbb E\bigl[(\Gap-\Rem)^{+}\bigr]$~\footnote{Here, $x^{+}:=\max\{x,0\}$ denotes the positive part of $x$.}.
\end{definition}

Carlsten {\em et al.}~\cite{carlsten2016instability} use an exponential model where $\Rem$ and $\Gap$ are i.i.d.\ exponential with mean $\mu$; this is the special case $\eta_R=\eta_G=\tfrac34\mu$, $p_G=\tfrac12$, and $\Delta=\tfrac{\mu}{2}$.

\begin{definition}[Scheme and subsidy primitives]
\label{def:scheme-primitives}
Fix a scheme $(\alpha,\beta,\delta)$ with $\alpha+\beta>0$ and a smoothing window $L\ge 1$. In round $t$, the realized fee mass is $F_t$; the canonical winner $W_t$ receives $\alpha F_t$, eligible non-winning atomic miners share $\beta F_t$ under the even or proportional rule (Section~\ref{subsec:redistribution-chi}), and $\delta F_t$ is deferred. The released smoothing amount at round $t$ is $S_t:=\frac{\delta}{L}\sum_{r=1}^{L}F_{t-r}$, ignoring initialization effects; it is split between the winner and the eligible non-winning atomic miners in proportions $\lambda_W:=\alpha/(\alpha+\beta)$ and $\lambda_L:=\beta/(\alpha+\beta)$. For the focal miner $i$, define the conditional subsidy means
\[
d_i:=\mathbb E\bigl[\text{current-fee loser payment}\bigm|\, i\neq W_t\bigr],\] \[s_i^W:=\mathbb E\bigl[\lambda_W S_t\bigm|\, i= W_t\bigr],\]
\[
d_i^{\mathrm{sm}}:=\mathbb E\bigl[\text{smoothing loser payment}\bigm|\, i\neq W_t\bigr],\]
\[d_i^{\mathrm{tot}}:=d_i+d_i^{\mathrm{sm}}.\]
\end{definition}

Finally, let $y\in[0,1]$ denote the total hash-power share, over all miners, that extends the default branch against a one-step fork. This includes Default-Compliant miners and, in our simulations, all non-atomic miners, since they are fixed to default behavior and excluded only from subsidy eligibility and learning updates. The complementary share $1-y$ resolves a one-step fork in favor of the branch that leaves more fees available, as in the Petty-Compliant tie-breaking rule of~\cite{carlsten2016instability}.

\begin{assumption}[One-step payoff proxy]
\label{assump:proxy}
Conditional on an opportunity for the focal atomic miner $i$, the local payoffs of the two actions are
\begin{equation}
\mathbb E[\mathrm{Base}_i\mid \mathrm{opp}]
= \alpha\,\mu_i^{\mathrm{opp}}+s_i^W, \label{eq:base-payoff}
\end{equation}
\begin{equation}
\mathbb E[\mathrm{FF}_i\mid \mathrm{opp}]
= \alpha\eta_R+(1-y)\,\alpha\eta_G+(1-y\,p_G)\,s_i^W+y\,p_G\,d_i^{\mathrm{tot}}. \label{eq:ff-payoff}
\end{equation}
\end{assumption}

Equations~\eqref{eq:base-payoff}--\eqref{eq:ff-payoff} rely on the following three idealizations, which we maintain throughout the analysis.

\begin{enumerate}
\item[(i)] \emph{One-step resolution.}
The next block resolves the fork race. When $\Rem>\Gap$, FF prescribes an ordinary extension that claims $\Rem$, yielding the contribution $\alpha\eta_R$. When $\Gap>\Rem$, FF prescribes an undercut. The undercutting fork becomes canonical if and only if the next miner is not default-following. This occurs with probability $1-y$ and yields the contribution $(1-y)\alpha\eta_G$.

\item[(ii)] \emph{Independence.}
The subsidy and smoothing quantities depend only on protocol state, including past fees and the identity of the winner. We assume that these quantities are independent of $(\Rem,\Gap)$. 

\item[(iii)] \emph{Current-round accounting.}
Payoffs include only rewards earned in the current round. 

\end{enumerate}

\begin{lemma}[Surplus identity]
\label{lem:surplus}
$\eta_R+\eta_G-\mu_i^{\mathrm{opp}}=\Delta\ge 0$, and $\Delta>0$ if and only if $p_G>0$.
\end{lemma}
\begin{proof}
Since $\Pr[\Rem=\Gap]=0$, we have $\Rem\,\mathbf 1\{\Rem>\Gap\}+\Gap\,\mathbf 1\{\Gap>\Rem\}=\max(\Rem,\Gap)$, so $\eta_R+\eta_G=\mathbb E[\max(\Rem,\Gap)]$. The identity follows from $\max(\Rem,\Gap)-\Rem=(\Gap-\Rem)^{+}$. Finally, $(\Gap-\Rem)^{+}$ is nonnegative and strictly positive exactly on $\{\Gap>\Rem\}$, so its expectation is positive iff $p_G>0$.
\end{proof}

\subsection{A General Threshold Theorem}
\label{subsec:threshold-theorem}

\begin{theorem}[Local undercutting threshold]
\label{thm:local-threshold}
Under Assumption~\ref{assump:proxy}, define the \emph{surplus term} and the \emph{deterrence margin}
\[
A_i:=\alpha\Delta\;\ge 0,
\qquad
B_i:=\alpha\eta_G+p_G\bigl(s_i^W-d_i^{\mathrm{tot}}\bigr).
\]
Then
\(
\mathbb E[\mathrm{FF}_i\mid \mathrm{opp}]-\mathbb E[\mathrm{Base}_i\mid \mathrm{opp}]=A_i-y\,B_i,
\)
so the one-step deviation is weakly profitable if and only if $y\,B_i\le A_i$. Consequently:
\begin{enumerate}
\item[(i)] If $B_i>0$, the deviation is weakly profitable precisely for $y\le y_i^{*}:=A_i/B_i$; a default-following share deters it iff $y>y_i^{*}$, and if $y_i^{*}\ge 1$ no feasible share $y\in[0,1]$ deters it.
\item[(ii)] If $B_i\le 0$, the deviation is weakly profitable for every $y\in[0,1]$, and its payoff advantage $A_i-yB_i$ is nondecreasing in $y$: additional default-following hash power does not deter, and can strictly encourage, the deviation.
\end{enumerate}
\end{theorem}
\begin{proof}
Subtracting \eqref{eq:base-payoff} from \eqref{eq:ff-payoff} and collecting the terms multiplying $y$ gives
\[
\alpha\bigl(\eta_R+\eta_G-\mu_i^{\mathrm{opp}}\bigr)
-y\Bigl(\alpha\eta_G+p_G\bigl(s_i^W-d_i^{\mathrm{tot}}\bigr)\Bigr),
\]
which equals $A_i-yB_i$ by Lemma~\ref{lem:surplus}. Both claims then follow from $A_i\ge0$: when $B_i>0$, the inequality $yB_i\le A_i$ reads $y\le A_i/B_i$; when $B_i\le 0$, it holds for every $y\ge0$, and $-B_i\ge 0$ makes the advantage nondecreasing in $y$.
\end{proof}

A failed undercut, which occurs with probability $y\,p_G$, forfeits the winner-side fee payment and the winner-side smoothing payment, while collecting the loser-side subsidy. Thus $B_i$ is exactly the rate at which default-following power penalizes the deviation: the loser-side subsidy always weakens deterrence, while the winner-side smoothing payment always strengthens it. This single expression organizes the remainder of the section.

\begin{corollary}[Winner-take-all]
\label{cor:wta}
Let $\alpha=1$, $\beta=\delta=0$, and $p_G>0$. Then $B_i=\eta_G>0$ and
\(
y^{*}_{\mathrm{wta}}:=\Delta/\eta_G
\)
for any joint fee distribution. If moreover $\Rem$ and $\Gap$ are i.i.d., continuous, and strictly positive with mean $\mu$, then, writing $\rho:=\mathbb E[\min(\Rem,\Gap)]/\mu\in(0,1)$,
\[
y^{*}_{\mathrm{wta}}
=2-\frac{2\mu}{\mathbb E[\max(\Rem,\Gap)]}
=\frac{2(1-\rho)}{2-\rho}\;\in(0,1).
\]
In particular, in the i.i.d.\ model a deterring default-compliant share always exists, and it depends on the fee distribution only through $\rho$.
\end{corollary}
\begin{proof}
Special case of Theorem~\ref{thm:local-threshold}.
\end{proof}

\begin{example}[Fee-distribution sensitivity]
\label{ex:fee-distributions}
Corollary~\ref{cor:wta} specializes as follows.
\begin{itemize}
\item \emph{Exponential fees}: $\rho=\tfrac12$, so $y^{*}_{\mathrm{wta}}=\tfrac23$, recovering the threshold of Carlsten {\em et al.}~\cite{carlsten2016instability}.
\item \emph{Uniform fees} on $[0,b]$: $\rho=\tfrac23$, so $y^{*}_{\mathrm{wta}}=\tfrac12$.
\item \emph{Pareto fees} with tail index $a>1$: $\rho=\tfrac{2(a-1)}{2a-1}$, so $y^{*}_{\mathrm{wta}}=\tfrac1a$.
\end{itemize}
The cutoff increases as $\rho$ decreases: heavy-tailed fee backlogs push $y^{*}_{\mathrm{wta}}$ toward $1$, making undercutting nearly undeterrable, while light-tailed fees push it toward $0$.
\end{example}

\begin{corollary}[Deferral-only: an FRSC idealization]
\label{cor:deferral}\label{subsec:frsc-overview}
Let $\beta=0$, $\delta\in[0,1)$, and $\alpha=1-\delta$, and suppose fees are stationary with per-block mean $\bar\mu:=\mathbb E[F_t]$ and $S_t$ independent of the winner identity. Then $s_i^W=\delta\bar\mu$, $d_i^{\mathrm{tot}}=0$, and
\[
y^{*}(\delta)=\frac{(1-\delta)\,\Delta}{(1-\delta)\,\eta_G+p_G\,\delta\,\bar\mu},
\]
which is strictly decreasing in $\delta$, from $\Delta/\eta_G$ at $\delta=0$ to $0$ as $\delta\to1$. Under the exponential model with $\mu=\bar\mu$ normalized to $1$,
\(
y^{*}(\delta)=\frac{2(1-\delta)}{3-\delta}.
\)
\end{corollary}
\begin{proof}
Special case of Theorem~\ref{thm:local-threshold}(i).
\end{proof}

Corollary~\ref{cor:deferral} is an idealization of the FRSC mechanism of Budinsk\'y {\em et al.}~\cite{budinsky2023fee}, under which the deferred payout behaves in expectation like a moving average of recent fee mass. It is not a reproduction of their simulations, which report that the required Default-Compliant share drops from $2/3$ to roughly $0.3$ in their main configuration. The corollary shows that the qualitative effect is already present in the local model, and explains it: deferring reward mass monotonically lowers the one-step cutoff (e.g.\ $y^{*}(\tfrac12)=0.4$) because a failed undercut forfeits the deferred claim while gaining nothing on the loser side.

\subsection{Splitting}
\label{subsec:redistribution-chi}
We now instantiate the loser-side subsidy under threshold-based atomicity. Let $\mathcal M$ denote the set of all miners and $\mathcal A\subseteq\mathcal M$ the set of atomic miners: miner $j$ is atomic iff its hash power $h_j$ is at least the cutoff $\gamma$. Let $H_{\mathcal A}:=\sum_{j\in\mathcal A}h_j$. Only atomic miners are eligible for the redistributed subsidy, but the canonical winner $W_t\in\mathcal M$ need not be atomic. Under proportional splitting, the loser-side payment to a focal atomic miner $i\in\mathcal A$ in round $t$ is
\[
R^{\mathrm{lose,prop}}_{i,t}
=\beta F_t\,\frac{h_i}{H_{\mathcal A}-h_{W_t}\mathbf 1\{W_t\in\mathcal A\}}\,\mathbf 1\{i\neq W_t\},
\]
\[
d_i^{\mathrm{prop}}:=\mathbb E\bigl[R^{\mathrm{lose,prop}}_{i,t}\bigm|\, i\neq W_t\bigr],
\]
and under even splitting it is
\[
R^{\mathrm{lose,even}}_{i,t}
=\beta F_t\,\frac{1}{|\mathcal A|-\mathbf 1\{W_t\in\mathcal A\}}\,\mathbf 1\{i\neq W_t\},
\]
\[
d_i^{\mathrm{even}}:=\mathbb E\bigl[R^{\mathrm{lose,even}}_{i,t}\bigm|\, i\neq W_t\bigr].
\]
Throughout, $d_i$ denotes $d_i^{\mathrm{prop}}$ or $d_i^{\mathrm{even}}$ according to the rule in force.

\begin{proposition}[Splitting alone cannot lower the local cutoff]
\label{prop:no-local-gain}
Fix $\delta=0$, any $\alpha\in(0,1]$, and any loser-side rule with $d_i\ge 0$, and suppose $p_G>0$. If $B_i=\alpha\eta_G-p_Gd_i>0$, then
\[
y_i^{*}(\alpha)
=\frac{\alpha\Delta}{\alpha\eta_G-p_Gd_i}
\;\ge\;\frac{\Delta}{\eta_G}=y^{*}_{\mathrm{wta}},
\]
with equality iff $d_i=0$, in which case $y_i^{*}$ does not depend on $\alpha$. If $B_i\le 0$, no feasible share deters the deviation. Hence, for every $(\alpha,\beta)$, the splitting-only local cutoff is never below the winner-take-all cutoff.
\end{proposition}
\begin{proof}
For $B_i>0$, since $d_i\ge0$ we have $\alpha\eta_G-p_Gd_i\le\alpha\eta_G$ and therefore $y_i^{*}=\alpha\Delta/(\alpha\eta_G-p_Gd_i)\ge\alpha\Delta/(\alpha\eta_G)=\Delta/\eta_G$, with equality iff $d_i=0$; in that case the factor $\alpha$ cancels. The case $B_i\le0$ is Theorem~\ref{thm:local-threshold}(ii).
\end{proof}

Proposition~\ref{prop:no-local-gain} is a deliberate negative result. Within the one-step model, the winner share $\alpha$ merely rescales both sides of the comparison, while the loser-side subsidy strictly cheapens failed undercuts. Any benefit of splitting must therefore operate through channels outside this local bound; the simulations of Section~\ref{subsubsec:strat-sim} identify such a channel, namely a shift of learned behavior toward the least aggressive fork-capable variants.

\begin{proposition}[Subsidy-dominated regime]
\label{prop:subsidy-dominated}
If $p_Gd_i\ge\alpha\eta_G$ (equivalently, $B_i\le0$ when $\delta=0$), then the one-step deviation is weakly profitable for every $y\in[0,1]$, and its advantage is nondecreasing in $y$. In this regime, failed undercuts are subsidized rather than punished, and default-following hash power ceases to act as a deterrent at all.
\end{proposition}
\begin{proof}
Immediate from Theorem~\ref{thm:local-threshold}(ii) with $s_i^W=d_i^{\mathrm{sm}}=0$.
\end{proof}

To obtain explicit formulas, we use a symmetric population:

\begin{definition}[Symmetric environment]
\label{def:symmetric-env}
There are $m\ge2$ atomic miners of equal hash power comprising almost all hash power; the winner $W_t$ is uniform over the $m$ miners and independent of $(F_t,S_t,\Rem,\Gap)$; fees are stationary with $\mathbb E[F_t]=\mu$; and $(\Rem,\Gap)$ are i.i.d.\ exponential with mean $\mu$, normalized to $\mu=1$. In this environment, even and proportional splitting coincide, with
\(
d_i=\beta/(m-1).
\)
\end{definition}

\begin{corollary}[Symmetric splitting-only cutoff]
\label{cor:symmetric-splitting}
In the symmetric environment with $\delta=0$, $B_i>0$ iff $\alpha>\bar\alpha_m:=\frac{2}{3m-1}$, and for such $\alpha$
\[
y_i^{*}(\alpha)
=\frac{\alpha/2}{\;\frac{3}{4}\alpha-\frac{1-\alpha}{2(m-1)}\;},
\]
which exceeds $\tfrac23$ whenever $\beta>0$ and tends to $\tfrac23$ as $\alpha\to1$ or $m\to\infty$; monotonicity in $\alpha$ is the case $\delta=0$ of Theorem~\ref{thm:smoothing}(ii). 
\end{corollary}
\begin{proof}
Substitute $\eta_G=\tfrac34$, $p_G=\tfrac12$, $\Delta=\tfrac12$ and $d_i=\tfrac{1-\alpha}{m-1}$ into Theorem~\ref{thm:local-threshold} and Proposition~\ref{prop:no-local-gain}; the boundary solves $\tfrac34\alpha=\tfrac{1-\alpha}{2(m-1)}$, i.e.\ $\alpha(3m-1)=2$. 
\end{proof}

\subsection{Splitting with Smoothing}
\label{subsec:splitting-and-smoothing-analytical}
We now activate the smoothing component: the scheme is $(\alpha,\,1-\alpha-\delta,\,\delta)$ with $\delta\in[0,1-\alpha]$. As in Section~\ref{subsec:smoothing-extension}, the deferred $\delta$-portion of each reward is released evenly over the next $L$ canonical blocks. To avoid recursively re-deferring already-deferred rewards, the released amount $S_t$ is treated as a released, protocol-determined quantity that future miners cannot selectively exclude or undercut; it is allocated between the winner and the eligible non-winning atomic miners by the same winner-versus-loser split normalized over the non-deferred shares, $\lambda_W=\alpha/(1-\delta)$ and $\lambda_L=\beta/(1-\delta)$, with the loser-side smoothing payment using the same proportional or even eligibility denominators as in Section~\ref{subsec:redistribution-chi}, applied to $\lambda_LS_t$ in place of $\beta F_t$. Instantiating Theorem~\ref{thm:local-threshold}, the smoothed cutoff in the regular regime $B_i>0$ is
\[
y_i^{*}(\alpha,\delta,L)
=\frac{\alpha\Delta}{\alpha\eta_G+p_G\bigl(s_i^W-d_i^{\mathrm{tot}}\bigr)}.
\]
Relative to the unsmoothed margin at the same $\alpha$, three effects compete: failed undercuts now also forfeit the winner-side smoothing payment $s_i^W$; losers receive a smoothing subsidy $d_i^{\mathrm{sm}}$; and the current-fee subsidy shrinks because $\beta$ decreases from $1-\alpha$ to $1-\alpha-\delta$. In the symmetric environment, the net effect admits an exact expression and is always strictly positive.

\begin{theorem}[Smoothing strictly improves the local cutoff]
\label{thm:smoothing}
In the symmetric environment, fix $\alpha\in(0,1)$ and $\delta\in[0,1-\alpha]$. Then
\(
s_i^W=\frac{\alpha\delta}{1-\delta}
\)
and
\(
d_i^{\mathrm{tot}}=\frac{1-\alpha-\delta}{(1-\delta)(m-1)},
\)
and the deterrence margin satisfies the exact identity
\begin{equation}
\label{eq:smoothing-identity}
B_i(\alpha,\delta)=B_i(\alpha,0)+\frac{m}{2(m-1)}\cdot\frac{\alpha\,\delta}{1-\delta}.
\end{equation}
\end{theorem}
\begin{proof}
By independence and stationarity, $\mathbb E[S_t]=\delta$, so $s_i^W=\lambda_W\,\mathbb E[S_t]=\frac{\alpha\delta}{1-\delta}$. The loser-side smoothing payment divides $\lambda_LS_t$ among the $m-1$ non-winners, so $d_i^{\mathrm{sm}}=\frac{\beta\delta}{(1-\delta)(m-1)}$ with $\beta=1-\alpha-\delta$, and
\(
d_i^{\mathrm{tot}}
=\frac{\beta}{m-1}\bigl(1+\frac{\delta}{1-\delta}\bigr)
=\frac{\beta}{(1-\delta)(m-1)}.
\)
Hence, with $\eta_G=\tfrac34$ and $p_G=\tfrac12$,
\[
B_i(\alpha,\delta)
=\frac{3\alpha}{4}
+\frac{1}{2(1-\delta)}\Bigl(\alpha\delta-\frac{1-\alpha-\delta}{m-1}\Bigr).
\]
Subtracting $B_i(\alpha,0)=\frac{3\alpha}{4}-\frac{1-\alpha}{2(m-1)}$ over the common denominator $2(1-\delta)(m-1)$ leaves the numerator
\(
\alpha\delta(m-1)-(1-\alpha-\delta)+(1-\alpha)(1-\delta)
=\alpha\delta(m-1)+\alpha\delta
=\alpha\delta m,
\)
which gives \eqref{eq:smoothing-identity}. 
\end{proof}

Consequently, \(B_i(\alpha,\delta)\) is strictly increasing in \(\delta\), so whenever \(B_i>0\), the cutoff
\[
y_i^*(\alpha,\delta)
=
\frac{\alpha/2}{B_i(\alpha,\delta)}
\]
strictly decreases with \(\delta\); hence any \(\delta>0\) yields a lower cutoff than splitting alone. For fixed \(\delta\), \(y_i^*\) also strictly decreases with \(\alpha\) in the regular regime, with the local optimum attained at \(\beta=0\), recovering the deferral-only cutoff of Corollary~\ref{cor:deferral}. If \(\alpha\le\bar{\alpha}_m\), so that \(\delta=0\) yields no deterrence, deterrence is restored whenever
\[
\delta>\delta_0(\alpha,m)
:=
\frac{\left|B_i(\alpha,0)\right|}
{\frac{\alpha m}{2(m-1)}+\left|B_i(\alpha,0)\right|},
\]
and \(\delta_0<1-\alpha\) since
\(
B_i(\alpha,1-\alpha)
=
\frac{3}{4}\alpha+\frac{1-\alpha}{2}>0
\). The cutoff is independent of the window length \(L\), since only \(\mathbb{E}[S_t]=\delta\) enters the one-step comparison. Finally,
\[
y_i^*(\alpha,\delta)
\xrightarrow[m\to\infty]{}
\frac{2(1-\delta)}{3-\delta},
\]
independently of \(\alpha\); for example, \(\delta=0.2\) gives approximately \(0.571\), compared with the unsmoothed limit \(2/3\).


\subsection{Restricted-policy MILP proxy}
\label{subsec:restricted-milp-proxy}

Theorem~\ref{thm:local-threshold} and its corollaries analyze a single one-step Function-Fork deviation. As an additional diagnostic, one can also formulate a restricted equilibrium problem over a finite Function-Fork family. It serves as a potential analytical proxy for examining whether aggressive undercutting policies can be eliminated in a discretized local model. Our formulation follows closely with~\cite{wolsey2014integer, wolsey2020integer,roughgarden2010algorithmic}.

At a high level, fix a finite family of Function-Fork policies \(U=\{1,1/2,1/4,1/8\}\), where smaller values correspond to more aggressive undercutting, and let \(U_{\mathrm{aggr}}\subseteq U\) denote the aggressive subset. For each local blockchain state \(\omega\), the model precomputes the included fee mass if policy \(u\) becomes canonical, the loser-side subsidy if the focal miner loses, and the one-step extension indicators for different next-miner types. The MILP then asks whether the remaining strategic mass can be assigned only to non-aggressive policies while ensuring that no aggressive policy gives a strictly higher one-step payoff.

This restricted MILP, given in Appendix~\ref{app:milp}, should be interpreted only as a local incentive proxy. It is not the same object as the learning-based strategy distributions reported in our simulations: the MILP studies one-step restricted best responses, whereas the simulator tracks repeated no-regret learning over many games. 

We solve the unsmoothed MILP formulation for \(\alpha\in\{0.25,0.50,0.75\}\), under both even and proportional splitting. In the exported one-step coefficients, the loser-side subsidy, \(\rho_u(\omega)=0\), in the sampled states: under the local failure-branch proxy, a failed focal fork continues on the current published non-focal branch, which often has zero remaining fee mass. 

Across the tested unsmoothed parameter points, the conservative MILP requirement remains near \(y_{\mathrm{req}}\approx 0.40\). This is qualitatively consistent with the unsmoothed simulation threshold of approximately \(0.36\) in Section~\ref{subsubsec:default-threshold}, although the two quantities are not identical: the MILP is a one-step restricted-equilibrium proxy, whereas the simulator reports repeated-learning outcomes. The near-flat dependence on \(\alpha\) is also expected, since with \(\rho_u(\omega)=0\), changing \(\alpha\) mostly rescales the candidate winner-side payoffs.

\subsection{Interpretation of the Analytical Results}
\label{subsec:analysis-interpretation}

The results above are local profitability cutoffs for one-step Function-Fork deviations: a smaller cutoff $y_i^{*}$ is better, meaning that less Default-Compliant hash power suffices to make the deviation unprofitable. Three conclusions are proved within the local model.

First, the baseline cutoff is distribution-sensitive: $2/3$ in~\cite{carlsten2016instability} is the exponential instance of $y^{*}_{\mathrm{wta}}=2(1-\rho)/(2-\rho)$, and heavier-tailed fee distributions push the cutoff toward $1$. Reward mechanisms should therefore be judged by how they move the cutoff for a given fee distribution, not against $2/3$ as an absolute constant.

Second, the loser-side subsidy is locally a pure liability. Splitting alone never lowers the cutoff below the winner-take-all value, and once $p_Gd_i\ge\alpha\eta_G$ the deviation is profitable for every $y$, because failed undercuts are subsidized rather than punished. This delineates the parameter region where the mechanism must not be pushed: driving $\alpha$ very low makes the redistribution component dominant and destroys local deterrence, consistent with the unstructured learned behavior observed for small $\alpha$ in Section~\ref{subsubsec:strat-sim}.

Third, smoothing is locally a strict improvement. The winner-side smoothing payment is forfeited exactly when an undercut fails, so every unit of deferred reward raises the deterrence margin; the improvement is monotone in $\delta$, independent of the window length $L$ in the mean-based local model, and approaches the deferral-only cutoff $2(1-\delta)/(3-\delta)$ in large populations.

\begin{figure*}[t]
\centering
\begin{subfigure}[t]{0.40\textwidth}
    \centering
    \includegraphics[width=\textwidth]{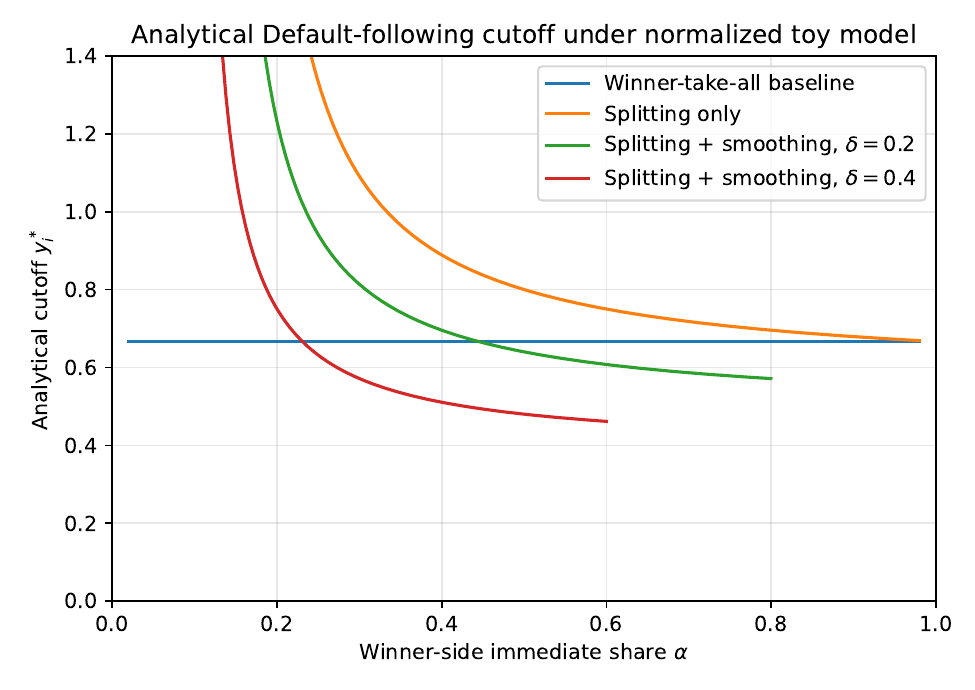}
    \caption{Feasible cutoff as a function of \(\alpha\)}
    \label{fig:analytical-bound-a}
\end{subfigure}
\hspace{0.05\textwidth}
\begin{subfigure}[t]{0.40\textwidth}
    \centering
    \includegraphics[width=\textwidth]{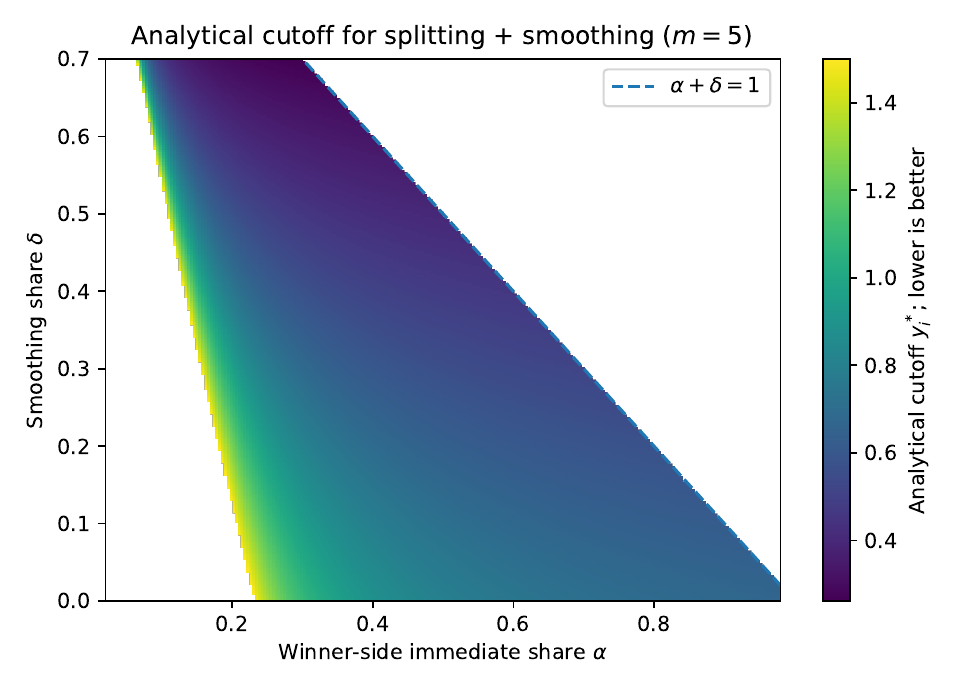}
    \caption{Regimes for splitting with smoothing}
    \label{fig:analytical-bound-b}
\end{subfigure}
\caption{Analytical behavior under the normalized exponential model.
Panel~(a) plots the cutoff \(y_i^*\) from
Theorem~\ref{thm:local-threshold}; only \(0\le y_i^*\le1\) is feasible,
while top markers denote \(y_i^*>1\).
Panel~(b) classifies infeasible, subsidy-dominated
(Proposition~\ref{prop:subsidy-dominated}), above-range, and feasible-cutoff regions.}
\label{fig:analytical-bound-regimes}
\end{figure*}

Figure~\ref{fig:analytical-bound-regimes} visualizes these regimes under the normalized exponential model. Panel~(a) shows that the splitting-only curve never falls below the $2/3$ baseline while smoothing shifts the feasible cutoff downward; panel~(b) separates the subsidy-dominated region of Proposition~\ref{prop:subsidy-dominated} from the regions with a meaningful feasible cutoff.

We summarize the relation of these results to the simulations results of the next section as follows. For threshold-based atomicity, the threshold experiments (Section~\ref{subsubsec:default-threshold}) find that splitting alone lowers the observed Default-Compliant threshold to $0.36$; by Proposition~\ref{prop:no-local-gain} this cannot be a one-step effect, and the strategy-learning experiments indeed attribute it to equilibrium selection: learned behavior shifts toward Function-Fork~$1$, the least aggressive fork-capable variant, rather than the aggressive undercutting that drives the $2/3$ bound. Smoothing, by contrast, improves both the local bound and the observed threshold (from $0.36$ to $0.23$), so the theory and the experiments move in the same direction, even though the magnitudes differ. The windowed atomicity mechanism, which we study only experimentally, lowers the observed threshold even further to approximately $0.19$. Finally, both levels of analysis point to an interior parameter region rather than an ``as much redistribution as possible'' rule: the local model penalizes small $\alpha$ through the subsidy-dominated regime, while $\alpha$ close to one approaches winner-take-all.

\section{Simulation}
\label{sec:simulation}
We complement our analysis with experiments performed using a mining simulator. We adapt the mining simulator produced by Carlsten {\em et al.} \cite{carlsten2016instability} to reflect our modified reward structure. In this section we describe the simulator modifications, then present the experiments used to evaluate the effect of our reward structure on strategy adoption, selfish mining profitability, and the Default-Compliant threshold needed to suppress profitable forking. 

\subsection{Simulator Overview and Modifications}
\label{subsec:sim-overview}

\subsubsection{Base Simulator}
\label{subsubsec:base-simulator}
The simulator of Carlsten {\em et al.} models repeated mining games in a transaction-fee-only regime. Each game proceeds over many rounds, where a round is the smallest unit of simulator time rather than a block. During these rounds, miners produce blocks probabilistically according to their hash power and receive rewards based on the blocks they contribute to the eventual longest chain. Mining strategies determine \textit{which} block a miner attempts to extend, \textit{how many} available transaction fees it includes, and \textit{when} newly found blocks are published. This allows the simulator to represent honest/default mining as well as fork-based deviations such as Petty Compliance, Function-Fork, and Lazy Fork strategies. 


The simulator also supports adaptive strategy selection through no-regret learning. In the learning experiments of Carlsten {\em et al.}, miners repeatedly choose from a fixed set of candidate mining strategies for the duration of a game. After each game, miner rewards are computed from the blocks they contributed to the longest chain, and the learning algorithm updates strategy weights for subsequent games, allowing the global strategy distribution to evolve over repeated play. In the baseline learning experiments directly comparable to ours, they use a simplified mining environment with $N$ miners of uniform hash power $1/N$.

We use this simulator as the basis for our experiments. Our implementation preserves the simulator's repeated-game framework, but modifies several components to reflect the reward-splitting mechanism proposed in Section~\ref{sec:modelPRS}. In particular, we change the reward calculation, add atomicity, implement even and proportional redistribution, incorporate optional smoothing, and study both synthetic and empirical mining distributions. We also modify the learning procedure: non-atomic miners are removed from the learning set as they are assumed to be default/honest, and learning updates are scaled by hash power so that updates reflect miner influence in heterogeneous mining environments.

Carlsten {\em et al.} also study selfish mining in the transaction-fee-only regime. We use the same underlying block simulation framework for our selfish mining experiments, but treat these experiments separately from the learning-based strategy simulations. In particular, our selfish mining experiments directly compare a single selfish miner against an honest baseline rather than updating strategy weights.

\subsubsection{Atomicity and Miner Distributions}
\label{subsubsec:sim-atomicity-distributions}
Since the proposed reward structure distributes subsidies only to eligible atomic miners, the simulator contains an explicit, hard-coded notion of atomicity. In experiments based on empirical mining distributions, each miner's hash-power share is compared against a fixed threshold $\gamma$. A miner is designated atomic if its share of global hash power is at least $\gamma$. In our experiments, we set $\gamma=0.01$, or $1\%$ of global hash power. Miners below this threshold are treated as non-atomic and are not eligible for the redistribution subsidy. 

The $1\%$ cutoff is a simple modeling decision rather than a claim of optimality. Moreover, in Bitcoin-esque PoW systems, a miner's current hash power share is not directly observable -- it can only be estimated from historical block production. In practice, atomicity would therefore need to be determined by observable proxies, e.g. whether a miner or pool has produced a block within a recent window of blocks. 

We consider both synthetic and empirical miner distributions. The synthetic environments use controlled miner populations, allowing us to vary quantities such as the fraction of Default-Compliant miners directly. These environment are useful for threshold experiments because the empirical mining distribution is fixed and does not permit controlled variation of the Default-Compliant fraction. 

For empirical experiments, we use three years of block-mining data from~\cite{blockchain_pool_distribution} to approximate a real-world mining-power distribution. Since a substantial portion of mined blocks are attributed to ``Unknown'' origin rather that to a known mining pool, we divide this unknown portion among $n$ identical non-atomic miners, to include the unknown hash power without treating it as a single strategic mining entity. The empirical distribution, shown in ~\ref{fig:empirical-mining-distribution}, has eight atomic miners with approximately $70\%$ of total hash power.


\subsubsection{Learning Updates}
\label{subsubsec:sim-learning-updates}
We additionally modify the learning procedure to better reflect heterogeneous mining power. In the original simulator, each miner is assigned one candidate mining strategy for the duration of a game. After the game completes, miner rewards are computed based on the blocks the contributed to the longest chain, and the no-regret learning algorithm updates strategy weights for the next game. 
This procedure is appropriate for a homogeneous mining environment in which miners have equal hash power, but in our heterogeneous environment however, miners may represent substantially different shares of the total mining power. We therefore scale learning updates by miner hash power, so that updates reflect each miner's influence on the mining process. Non-atomic miners are excluded from the learning set and remain Default-Compliant throughout the simulation.


\subsection{Simulation Experiments}
\label{subsec:sim-experiments}
We conduct four families of experiments with our modified simulator. First, we use the simulator's learning dynamics to study how the proposed reward structure affects strategy adoption among Petty-Compliant, Function Fork, and Lazy Fork strategies. Second, we run controlled Default-Compliant threshold experiments to estimate the minimum fixed Default-Compliant fraction at which Function-Fork strategies cease to hold a (combined) majority of the final learned strategy distribution. Third, we evaluate how splitting affects the profitability threshold for selfish mining. Fourth, we evaluate how the learned equilibrium strategies affect two observable measures of overall system health: canonical fee throughput and stale-block rate.

Our experiments measure different aspects of the proposed reward structure. The strategy learning experiments test whether splitting shifts learned behavior away from fork-based deviations. The Default-Compliant threshold experiments study how much honest/default participation is needed before Function-Fork strategies fall below majority learned weight. The selfish mining experiments use a direct revenue comparison, rather than learned strategy weights, to test whether selfish mining remains profitable under the modified reward rule. 

All simulation experiments use games with $10{,}000$ expected blocks and a target block interval of $600$ seconds. For the strategy-learning and Default-Compliant threshold experiments, each parameter point is run for $10{,}000$ games, after which we export the final learned strategy distribution. For the selfish mining experiments, each parameter point is evaluated over $10{,}000$ paired trials, where each trial runs both a selfish mining case and an analogous honest case for comparison.  

\subsubsection{Strategy-Learning}
\label{subsubsec:strat-sim}
\paragraph{Purpose.} 
The strategy-learning experiments evaluate whether splitting changes the strategies that strategic miners learn to favor. For each choice of reward splitting parameters, redistribution rule, and smoothing parameters, we observe the final learned strategy distribution produced by the learning-based strategy simulator. The central question is whether the modified reward structure shifts learned behavior away from fork-based deviations and toward less aggressive behavior such as Petty Honest/Petty Compliant mining. We note that the simulator provided by Carlsten {\em et al.} does not contain Honest/Default Compliant mining as the Petty variant is a clearly dominant strategy. These experiments therefore measure strategy adoption under repeated play, rather than the one-shot profitability of a single deviation. 

\paragraph{Setup.}
We run the strategy-learning experiments using the modified learning-based simulator described above. The learned strategy menu consist of Petty Honest/Petty Compliant mining, Function-Fork variants, and Lazy Fork. In the Function-Fork family, a miner may create a competing fork while choosing how much of the available transaction-fee pool to claim and how much to leave available to future miners. Following Carlsten {\em et al.}'s notation, larger Function-Fork indices correspond to more aggressive undercutting behavior: Function-Fork $1$ is the least aggressive variant, while Function-Fork $8$ leaves the most fees behind and is the most aggressive. Strictly speaking, Function-Fork $1$ is not an undercutting strategy in the fee-manipulation sense. It is better viewed as a fork-capable strategy that does not intentionally make its fork more attractive by leaving additional fees behind. Thus, larger Function-Fork indices represent forks that are more attractive to subsequent Petty Compliant miners, but less immediately rewarding to the undercutter if the fork succeeds.

We evaluate the learned strategy distribution under both even and proportional redistribution, with and without smoothing, across a sweep of reward-splitting parameters. For this experiment, miner hash power is derived from the historical mining pool data described in Section~\ref{subsubsec:sim-atomicity-distributions}. 
For each parameter point, we record the final learned strategy distribution after repeated play. All results for this experiment are averaged over ten independent trials. Each trial uses the same parameter settings, but an independent simulator run with figures reporting the mean value across these trials. For experiments with smoothing, a $\delta =0.2$ portion of each block reward was deferred and distributed across the following $L=5$ blocks. 

\paragraph{Results and Interpretation}
\begin{figure}[ht]
\centering

\begin{subfigure}[t]{0.48\textwidth}
    \centering
    \includegraphics[width=\textwidth]{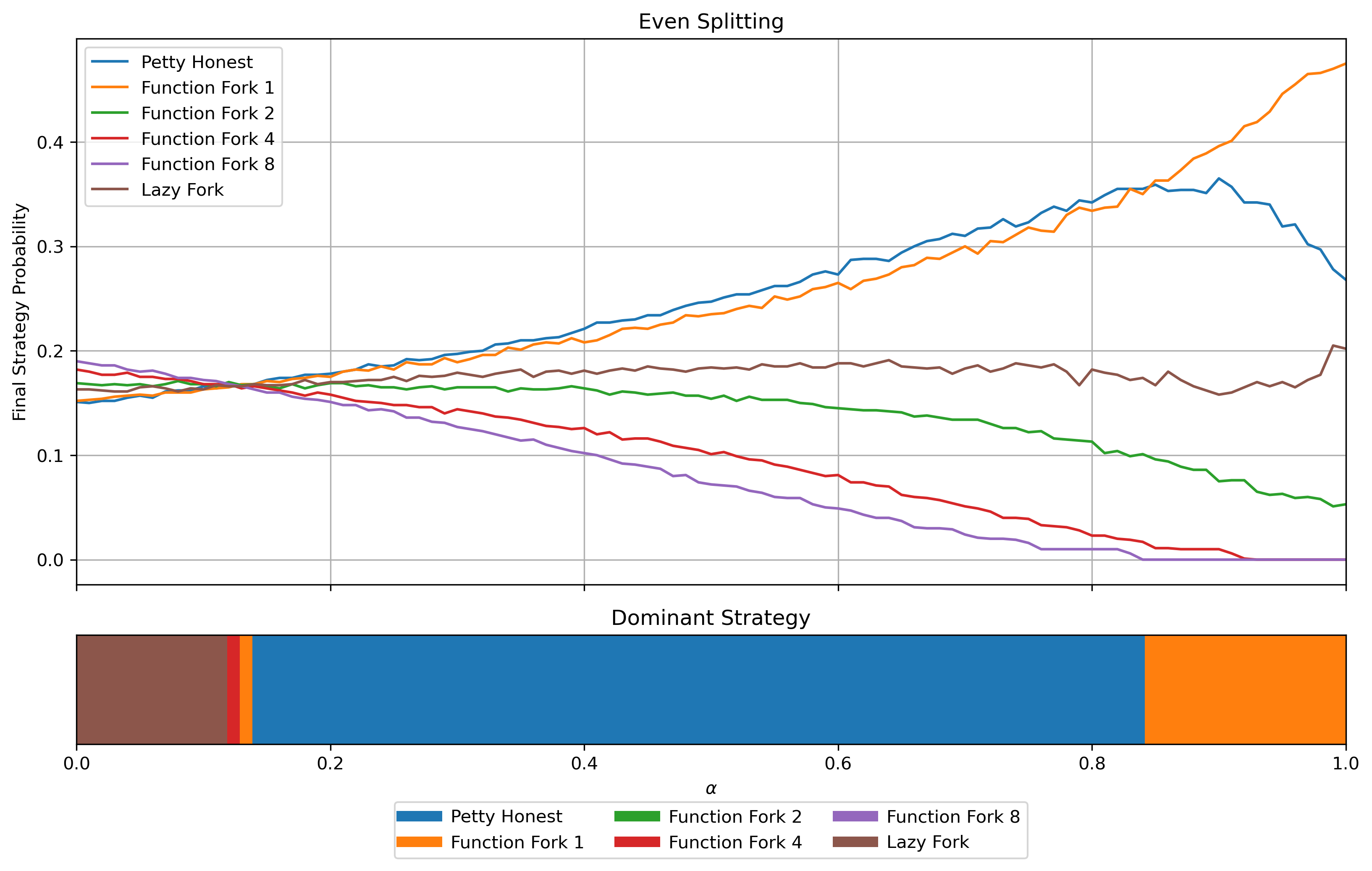}
    \caption{Even splitting, no smoothing.}
    \label{fig:strategy-learning-a}
\end{subfigure}
\hfill
\begin{subfigure}[t]{0.48\textwidth}
    \centering
    \includegraphics[width=\textwidth]{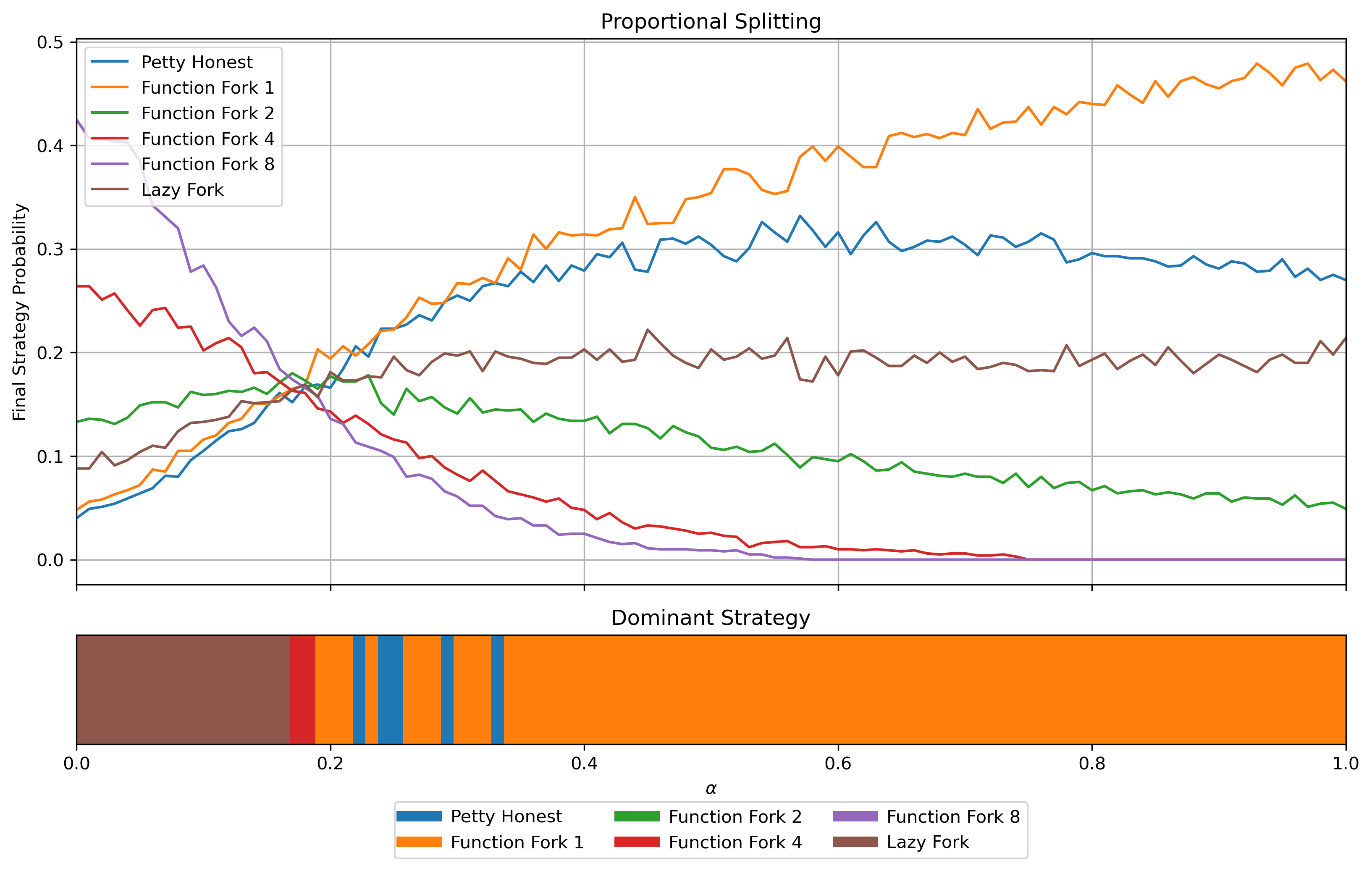}
    \caption{Proportional splitting, no smoothing.}
    \label{fig:strategy-learning-b}
\end{subfigure}

\vspace{0.75em}

\begin{subfigure}[t]{0.48\textwidth}
    \centering
    \includegraphics[width=\textwidth]{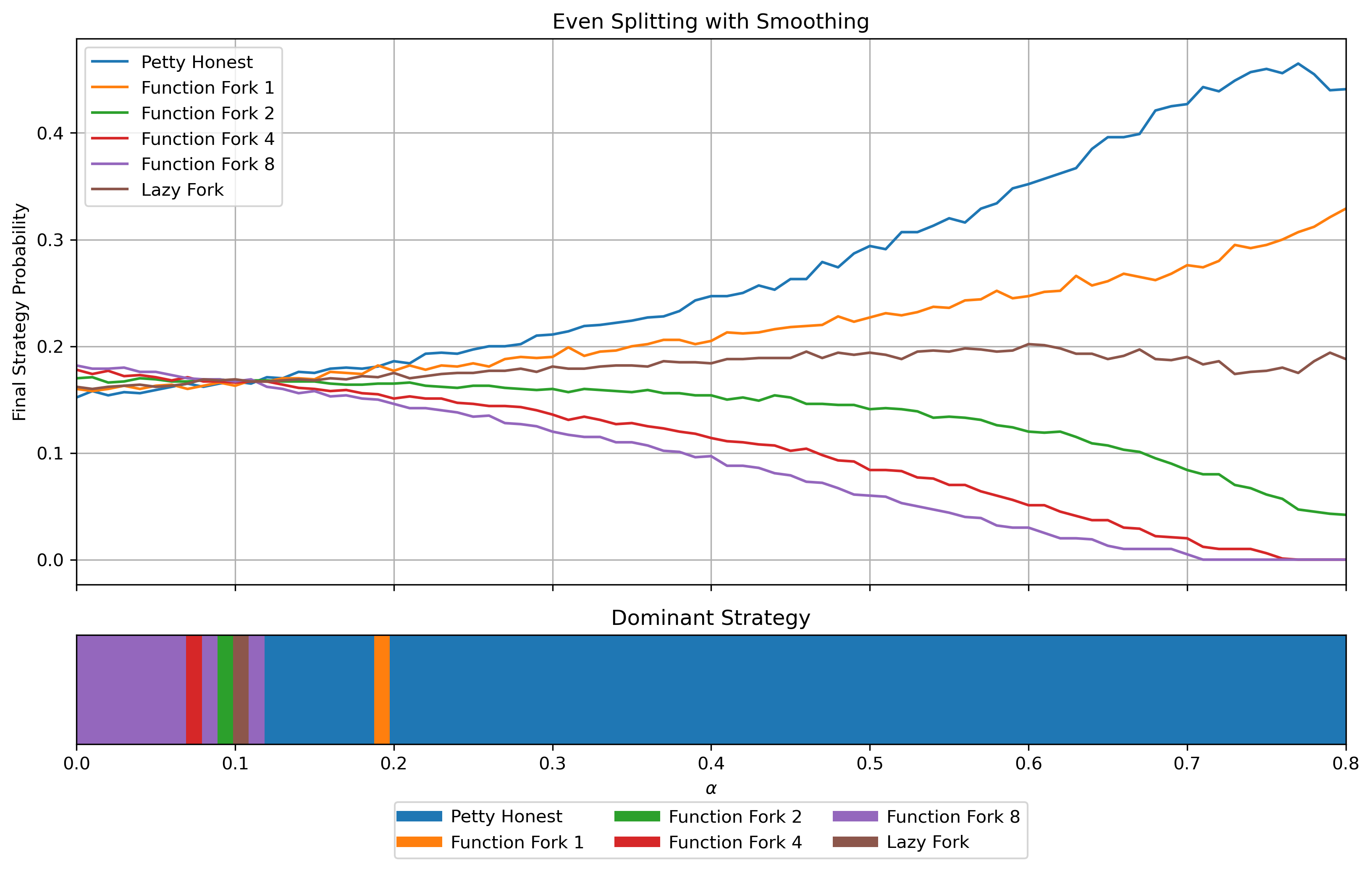}
    \caption{Even splitting with smoothing.}
    \label{fig:strategy-learning-c}
\end{subfigure}
\hfill
\begin{subfigure}[t]{0.48\textwidth}
    \centering
    \includegraphics[width=\textwidth]{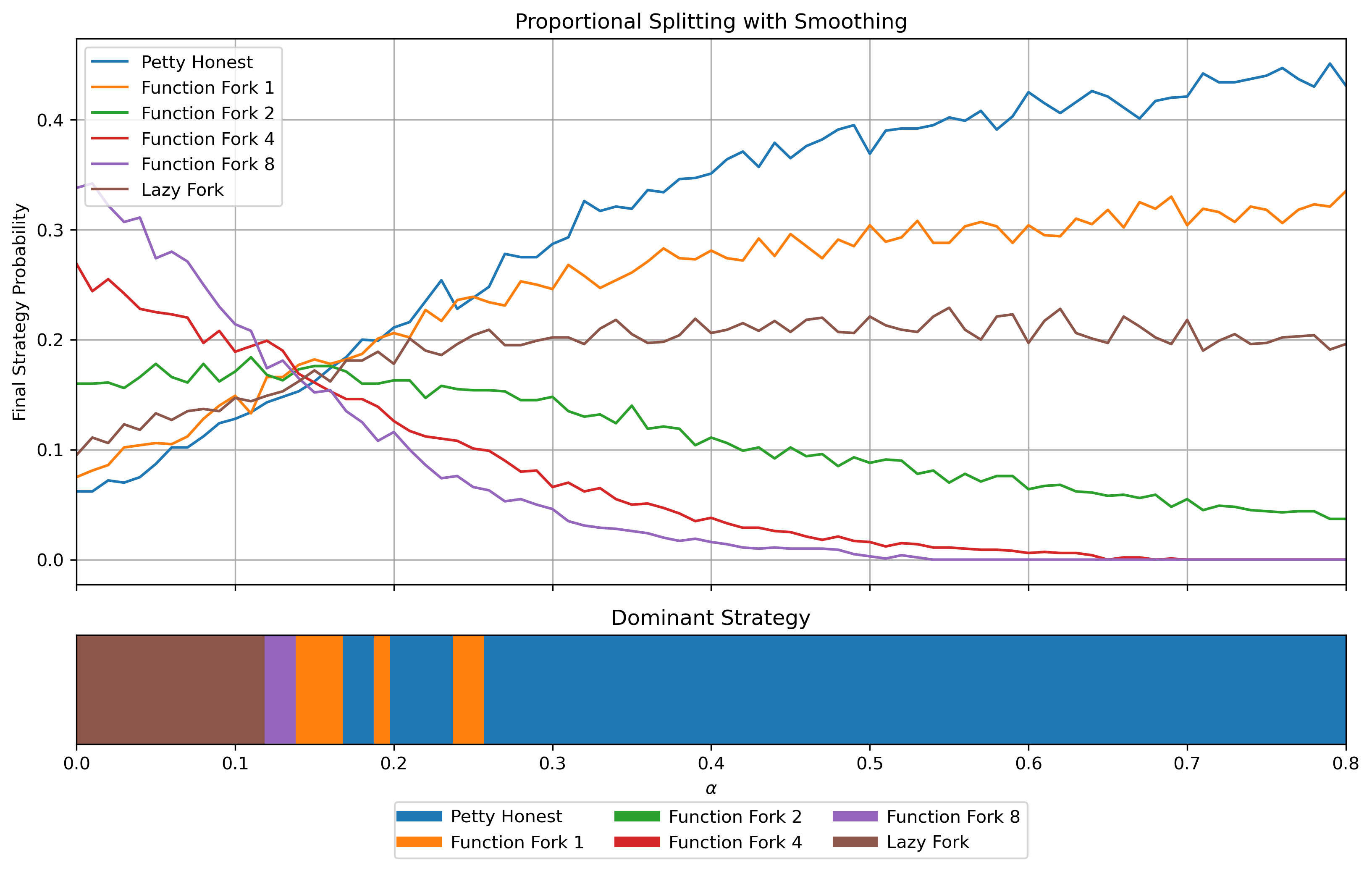}
    \caption{Proportional splitting with smoothing.}
    \label{fig:strategy-learning-d}
\end{subfigure}%

\caption{Strategy learning results under the proposed reward structure using the empirical mining pool distribution. Each panel shows the final learned strategy probabilities as a function of the winner-side reward fraction $\alpha$, averaged over ten independent trials. The lower bar indicates the dominant strategy at each value of $\alpha$. Smoothing panels use $\delta=0.2$ and a smoothing window $L=5$.}%
\label{fig:strategy-learning-panel}%
\end{figure}

The results, shown in Figure~\ref{fig:strategy-learning-panel}, suggest that splitting can substantially alter the strategy landscape observed by Carlsten {\em et al.}~\cite{carlsten2016instability}. In the original transaction-fee-only setting, the winner-take-all reward rule creates incentives for miners to adopt increasingly aggressive undercutting strategies, which yield greater rewards for strategic miners in the induced fork. 

We note that not every value of $\alpha$ in this sweep is realistic. For example, when $\alpha$ is very low, an atomic miner is likely incentivized to \textit{lose}, as they may capture a larger return via the $\beta$-subsidy. As can be seen in panels~\ref{fig:strategy-learning-a} and~\ref{fig:strategy-learning-c}, there is not a clear separation between the strategies for (roughly) $\alpha \in [0.0, 0.2]$. We posit that the reason for this is that the choice of mining strategies is largely meaningless due to the $\beta$-subsidy providing a majority of miner income. Additionally, as $\alpha$ nears $1.0$ the system approaches a winner-take-all reward structure. However, we note that only under even splitting without smoothing (panel \ref{fig:strategy-learning-a}) does a forking strategy overtake an honest one. However, the dominant forking strategy is the \textit{least} aggressive among those available to the miners, providing greater transaction throughput.  From these observations, we infer that $\alpha \in [0.2,0. 8]$ to be the most promising in achieving a system where honest strategies dominate deviant ones. 

Another interesting result may be seen in \ref{fig:strategy-learning-b}. Under proportional splitting without smoothing, Function-Fork $1$ dominates across nearly the entire range of $\alpha$. This should not be interpreted as a return to the aggressive undercutting behavior identified by Carlsten {\em et al.}~\cite{carlsten2016instability}. Function-Fork $1$ is the least aggressive Function-Fork variant, and unlike the higher index Function-Fork strategies, it does not leave a large fee surplus to attract Petty Compliant miners. One possible explanation is that proportional redistribution subsidizes larger non-winning atomic miners more heavily, reducing their downside risk of forking behavior. Since Function-Fork $1$ does not intentionally leave a large fee surplus behind, it avoids the immediate reward sacrifice associated with more aggressive variants while retaining the option to compete on a fork. 
This is still a meaningful improvement over the undercutting behavior identified by Carlsten {\em et al.} Thus, even in the parameter region where it dominates, the reward mechanism appears to reduce the impact on blockchain functionality.

\subsubsection{Default-Compliant Threshold}
\label{subsubsec:default-threshold}

\paragraph{Goal and Setup.}
The Default-Compliant threshold experiments estimate how much honest/default participation is needed before fork-based strategies cease to dominate the learned strategy distribution. In order to directly vary the fraction of miners committed to Default-Compliant behavior, we use a controlled synthetic environment in which we sweep the Default-Compliant miner fraction and determine, for each $\alpha$, how much Default-Compliant participation is sufficient to suppress Function-Fork Strategies.
This environment contains $100$ miners of equal hash power (notably, \textit{all} miners are then atomic). For a given value Default-Compliant fraction $c$, a $c$-fraction of the synthetic miners are assigned to the honest strategy throughout the duration of the experiment. The remaining fraction of miners participate in the learning process over the strategy set. This allows us to directly measure what fraction of Default-Compliant miners is needed to suppress Function-Fork strategies. 

For each reward environment, we sweep the winner-side reward fraction $\alpha$ and the Default-Compliant fraction $c$. For smoothed experiments, we stay consistent with previous parameters $\delta=0.2$, $L=5$. At each parameter point, we run the learning simulator and record the final learned strategy distribution.

\begin{figure}[ht]
\centering

\begin{subfigure}[t]{0.48\textwidth}
    \centering
    \includegraphics[width=\textwidth]{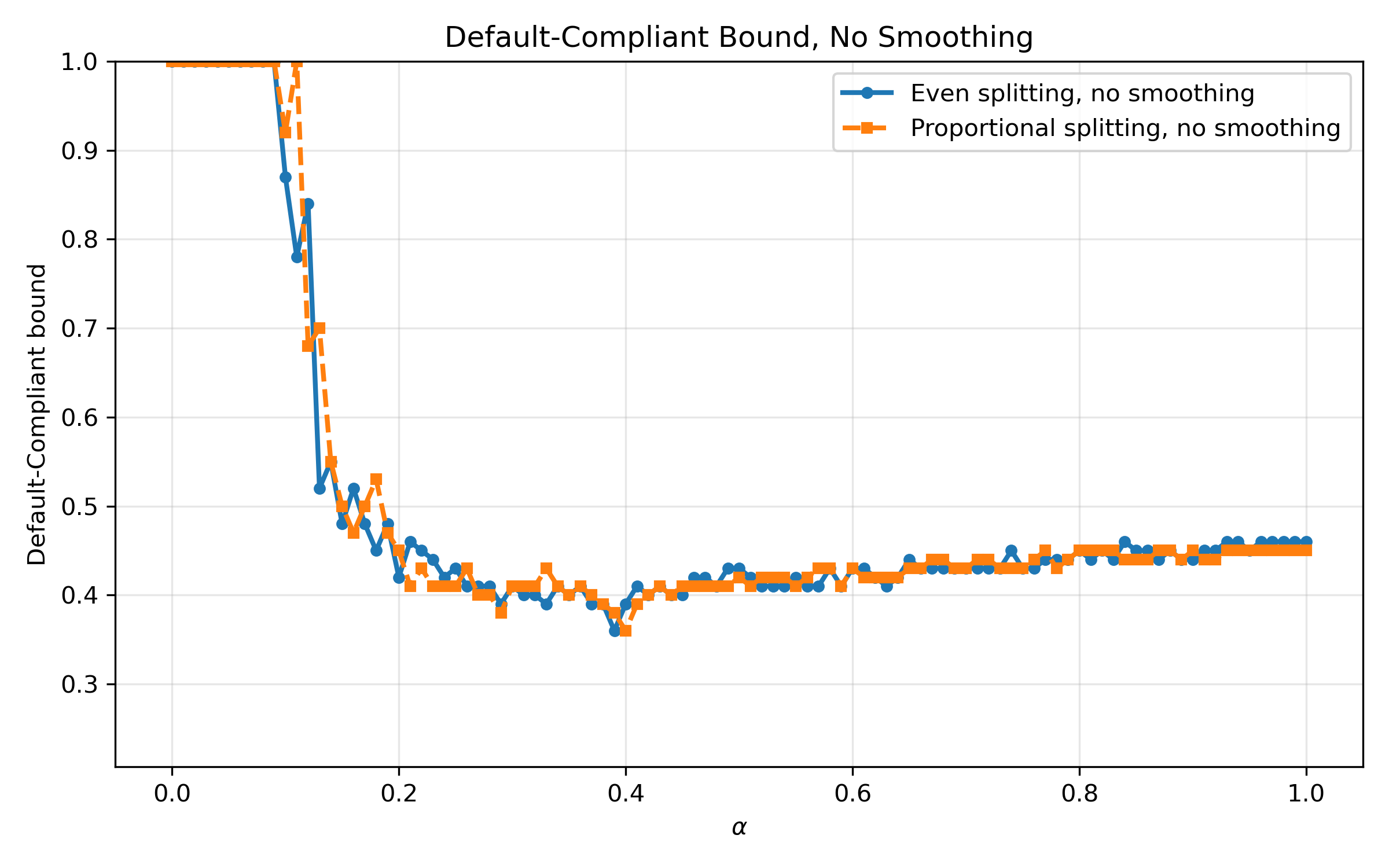}
    \caption{No smoothing}
    \label{fig:default-threshold-a}
\end{subfigure}
\hfill
\begin{subfigure}[t]{0.48\textwidth}
    \centering
    \includegraphics[width=\textwidth]{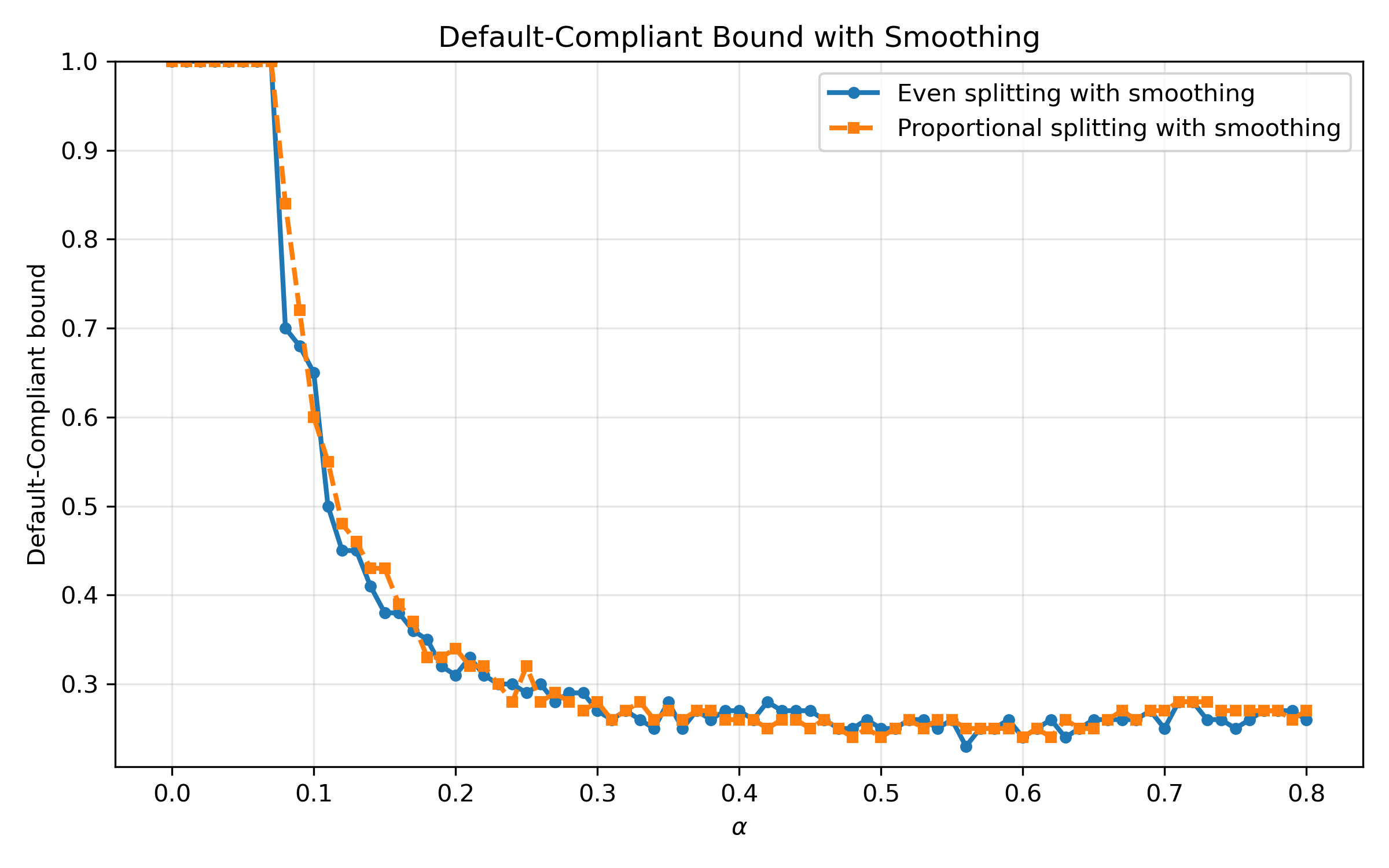}
    \caption{Smoothing}
    \label{fig:default-threshold-b}
\end{subfigure}

\vspace{0.75em}

{\small
\begin{tabular}{lcc}
\hline
Reward environment & Min. Default-Compliant bound & $\alpha$ value(s) \\
\hline
Even splitting, no smoothing
    & $0.36$
    & $0.39$ \\
Proportional splitting, no smoothing
    & $0.36$
    & $0.40$ \\
Even splitting with smoothing
    & $0.23$
    & $0.56$ \\
Proportional splitting with smoothing
    & $0.24$
    & $0.48,\;0.50,\;0.60,\;0.62$ \\
\hline
\end{tabular}
}

\caption{Default-Compliant threshold results. Each panel reports, as a function of $\alpha$, the smallest Default-Compliant fraction at which the combined learned probability mass on Function-Fork strategies falls below $0.5$. 
Due to the computational intensity, we report the average of two trials.}
\label{fig:default-threshold-panel}
\end{figure}

\paragraph{Results and Interpretation.}
The threshold experiment results, shown in Figure~\ref{fig:default-threshold-panel}, show that the proposed reward mechanism can substantially reduce the amount of fixed Default-Compliant participation required to suppress Function-Fork behavior. In the unsmoothed environments, the best observed threshold is $36\%$, already well below the $66\%$ baseline identified by Carlsten {\em et al.}~\cite{carlsten2016instability}. Incorporating smoothing further reduces the observed threshold to approximately $23\%$, below the approximately $30\%$ identified by Budinsk\'y {\em et al.} for their FRSC mechanism~\cite{budinsky2023fee}. These results indicate that the immediate redistribution among eligible atomic miners is beneficial on its own, and that smoothing provides an additional stabilizing effect.

\subsubsection{Selfish-Mining Threshold}
\label{subsubsec:selfish-mining}
\paragraph{Purpose and setup.}
Selfish mining is a well known and thoroughly studied mining attack in Bitcoin and similar PoW systems \cite{eyal2014majority} \cite{sapirshtein2016optimal} \cite{nayak2016stubborn}. In selfish mining, an attacker withholds newly mined blocks to build a private chain. The attacker then selectively publishes blocks from the private chain to either force public tied states, or should their lead become too short, publish their entire private chain; claiming all rewards by the longest chain rule. 

Selfish mining is typically studied by determining the computational power required for the attacker to be profitable. Making use of the selfish mining component of Carlsten {\em et al.}'s simulator, we perform the same style of threshold study under our reward scheme. That is, for each reward environment, we compare the revenue share of a single selfish miner against an analogous honest baseline and identify the attacker hash power at which selfish mining becomes profitable. 
For each parameter point, we run paired simulations. In one simulation, the focal miner follows the selfish strategy; in the other, the same focal miner follows the honest strategy. We then compare the focal miner's revenue share in the selfish case to its revenue in the honest baseline.


This paired comparison is necessary because the classical selfish mining profitability criterion, which compares an attacker's revenue share to its hash power share, is not well defined for our reward structure. Under splitting, an atomic miner may receive subsidies even when it does not mine successfully, so its honest revenue share need not equal its computational share. We therefore define profitability relative to the honest baseline under the same reward rule: selfish mining is profitable only when the focal miner earns more by following the selfish strategy than it would by mining honestly.

The selfish mining threshold experiments use a controlled synthetic mining distribution rather than the empirical one. Each simulation contains a focal atomic miner whose whose hash power is swept across the tested range, a non-atomic ocean miner with fixed at $25\%$, and five uniform atomic miners that split the remainder. 
We evaluate this profitability threshold across the same reward environments considered above: even/proportional splitting, and their smoothing variants (with $\delta=0.2$, $L=5$). In each case, we sweep the focal miner's hash power share and record the smallest share at which selfish mining is profitable. 

Selfish mining profitability also depends on how ties are resolved when the attacker publishes a private block at the same height as the public chain. In the standard selfish mining literature, this is typically denoted by $\gamma$, the fraction of honest mining power that mines on the attacker's block during a tie ~\cite{eyal2014majority, sapirshtein2016optimal}. As $\gamma$ is already used in this work for the atomicity threshold, we denote the selfish mining tie-breaking or "influence" parameter by $\tau$. We consider two settings: $\tau=0$ and $\tau=1/2$. These values corresponding to the attacker having no tie breaking advantage, and where half the honest mining power mines on the attacker's block during a tie, respectively. These settings are not meant to imply that honest miners \textit{intentionally} violate the protocol, but rather model different levels of advantage an attacker \textit{may} have in block propagation races.

\begin{figure}[ht]
\centering

\begin{subfigure}[t]{0.48\textwidth}
    \centering
    \includegraphics[width=\textwidth]{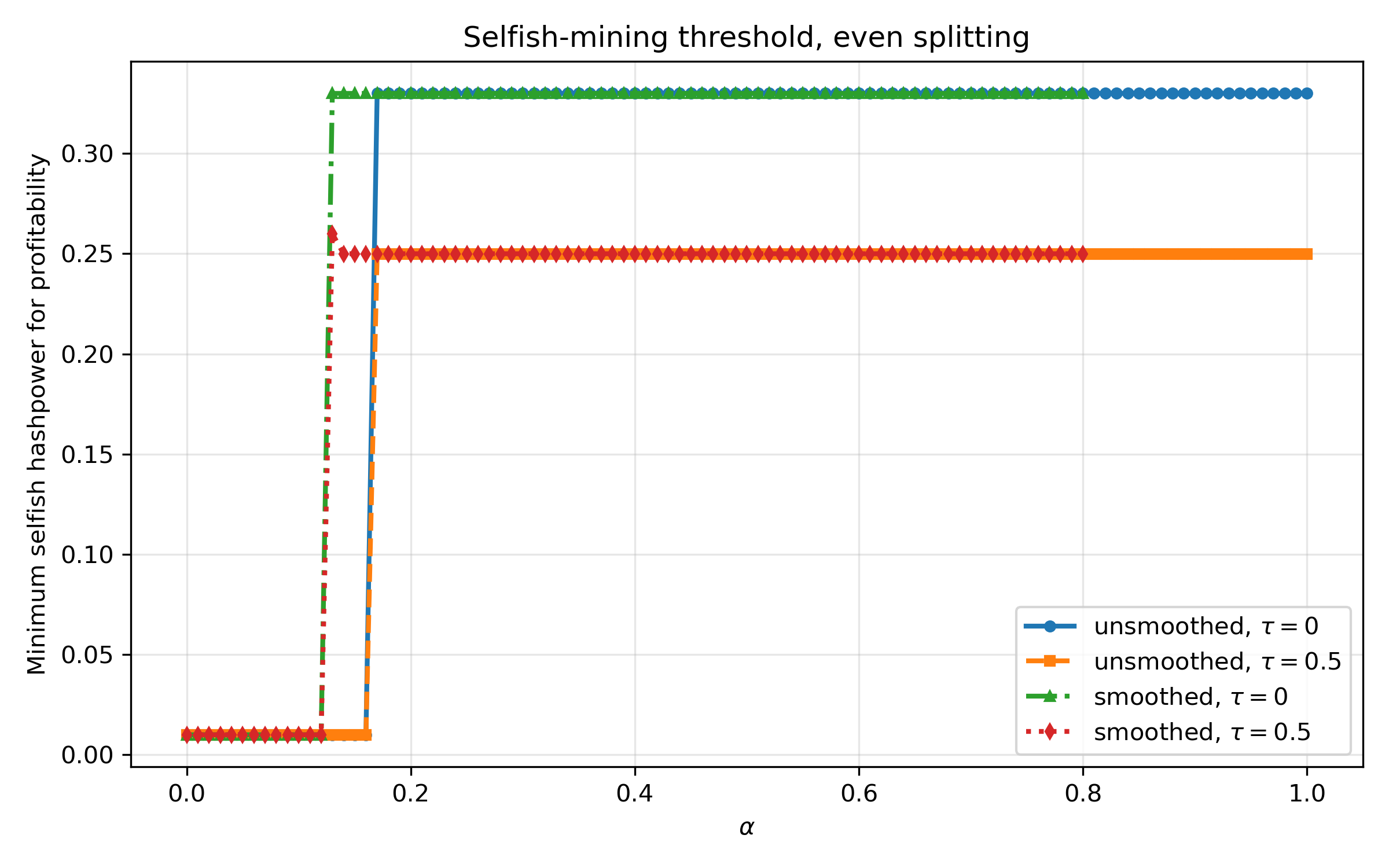}
    \caption{Even splitting}
    \label{fig:selfish-even}
\end{subfigure}
\hfill
\begin{subfigure}[t]{0.48\textwidth}
    \centering
    \includegraphics[width=\textwidth]{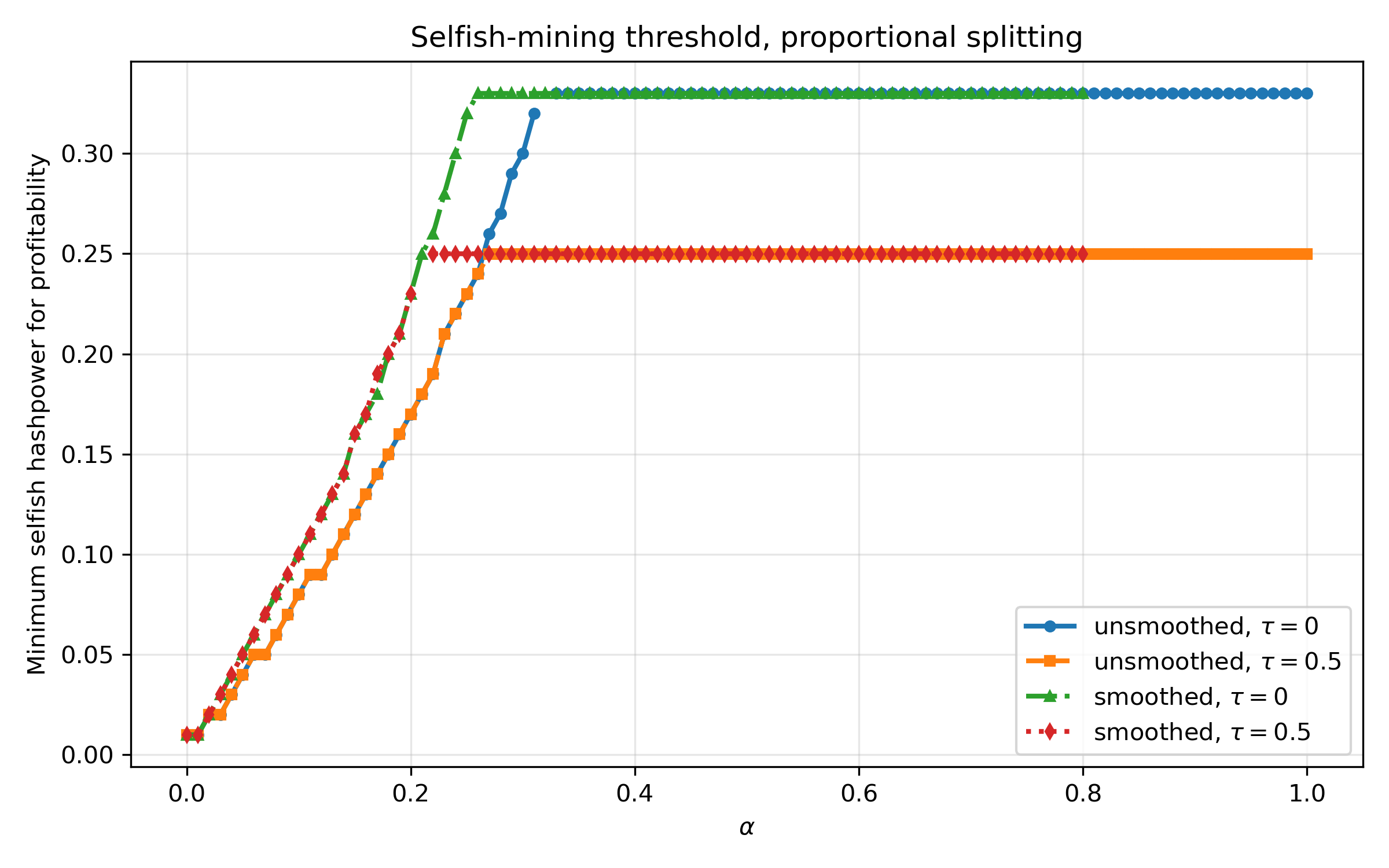}
    \caption{Proportional splitting}
    \label{fig:selfish-prop}
\end{subfigure}

\vspace{0.75em}


\caption{Selfish mining threshold results under the proposed reward structure. Each panel shows the minimum bound for profitability of selfish mining; that is, when selfish mining earns more in expectation than honest mining would.}
\label{fig:selfish-threshold-panel}
\end{figure}

\paragraph{Results and Interpretation.}
The results are shown in Figure~\ref{fig:selfish-threshold-panel}.  Our main observation is that the proposed reward mechanism does not make selfish mining easier than in the winner-take-all setting. In all environments, the selfish mining profitability threshold reaches and flatlines at the classical bound: approximately $1/3$ when $\tau=0$, and approximately $1/4$ when $\tau=1/2$. These values are consistent with the theoretical bounds identified by ~\cite{eyal2014majority}. The important behavior is not the case where $\alpha$ is close to $1$, since that regime is essentially the standard winner-take-all setting. Rather, the key observation is that for a broad range of nontrivial reward splits, the mechanism does not lower the selfish mining threshold below the classical bound. Instead, as the winner-side reward becomes large enough for block winning incentives to dominate the redistribution subsidy, the threshold rises to the standard bound and stays there. 

This is important because Carlsten {\em et al.} show that selfish mining behaves differently in the transaction-fee-only regime: it performs slightly better under transaction fees than under fixed block rewards. They also introduce stronger selfish mining variants. Our experiments do not study those variants, instead they ask whether our reward split worsens the profitability of the simulator's (orthodox) selfish mining strategy. The results suggest that it does not.

\subsubsection{System Health}
\paragraph{Goals and setup.}
The strategy-learning results identify which strategies are learned at equilibrium, but they do not themselves measure the system-level consequences of those strategies. In particular, Function-Fork $1$ is fork-capable but does not intentionally undercut by leaving additional fees behind. We therefore evaluate two observable system-health metrics: canonical fee throughput and stale-block rate. Canonical fee throughput is the fraction of generated fee mass that ultimately appears in the canonical chain, defined as 
$\text{canonical fee mass}/\text{total generated fee mass}$. Stale-block rate is the fraction of mined blocks that do not ultimately appear in the canonical chain, defined as 
$(\text{total mined blocks} -\text{canonical blocks}) / \text{total mined blocks}$.  Together these metrics separate fee-inclusion effects from wasted-work effects. 

For each reward environment and each value of $\alpha$, we take the final learned strategy distribution from the strategy-learning experiments~\ref{subsubsec:sim-learning-updates}. We then run fixed-strategy evaluation simulations: miners sample strategies according to the learned distribution for that value of $\alpha$, but no further learning occurs. 

\begin{figure}[tp]
\centering

\begin{subfigure}[t]{0.48\textwidth}
    \centering
    \includegraphics[width=\textwidth]{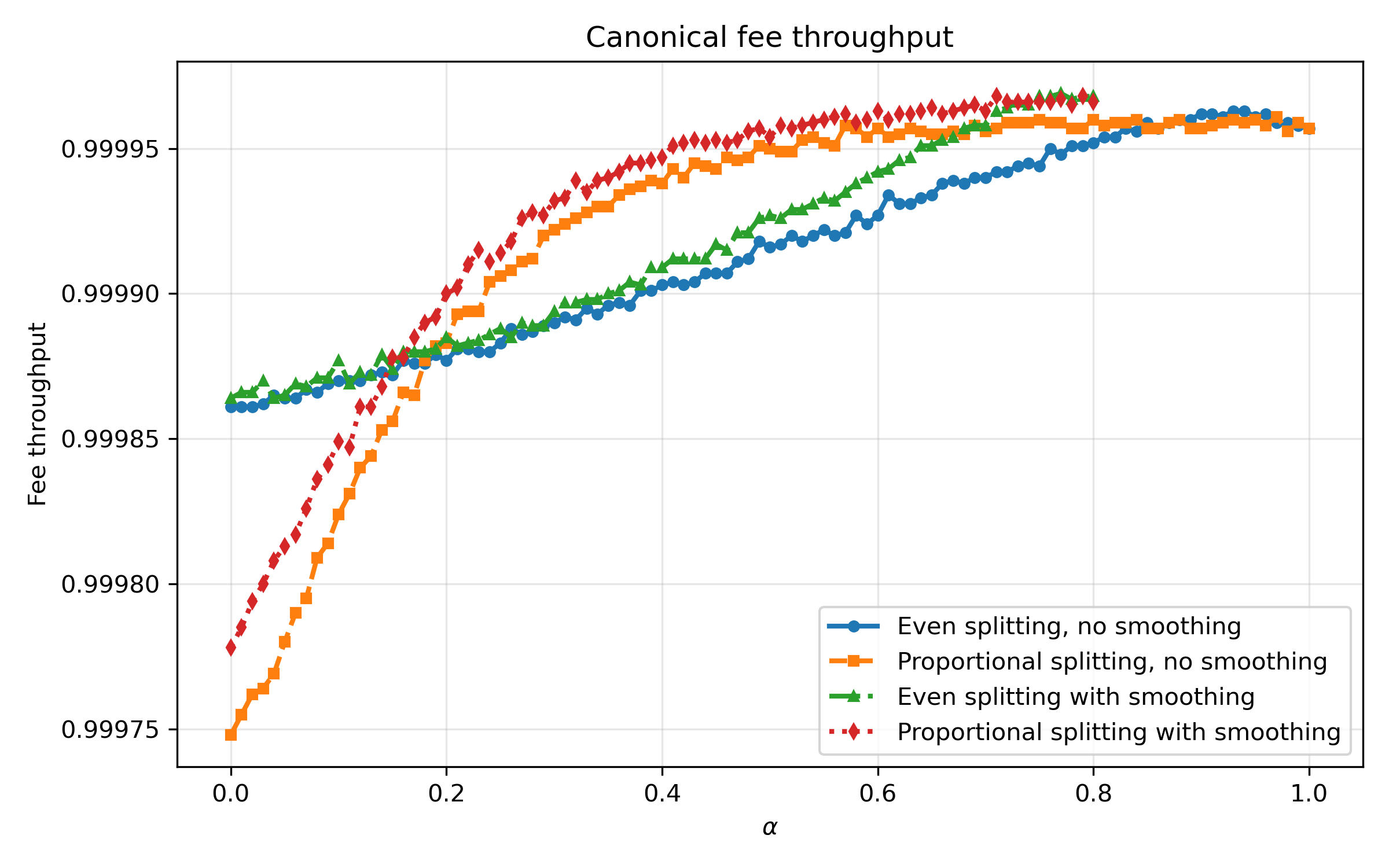}
    \caption{Fee throughput}
    \label{fig:throughput-a}
\end{subfigure}
\hfill
\begin{subfigure}[t]{0.48\textwidth}
    \centering
    \includegraphics[width=\textwidth]{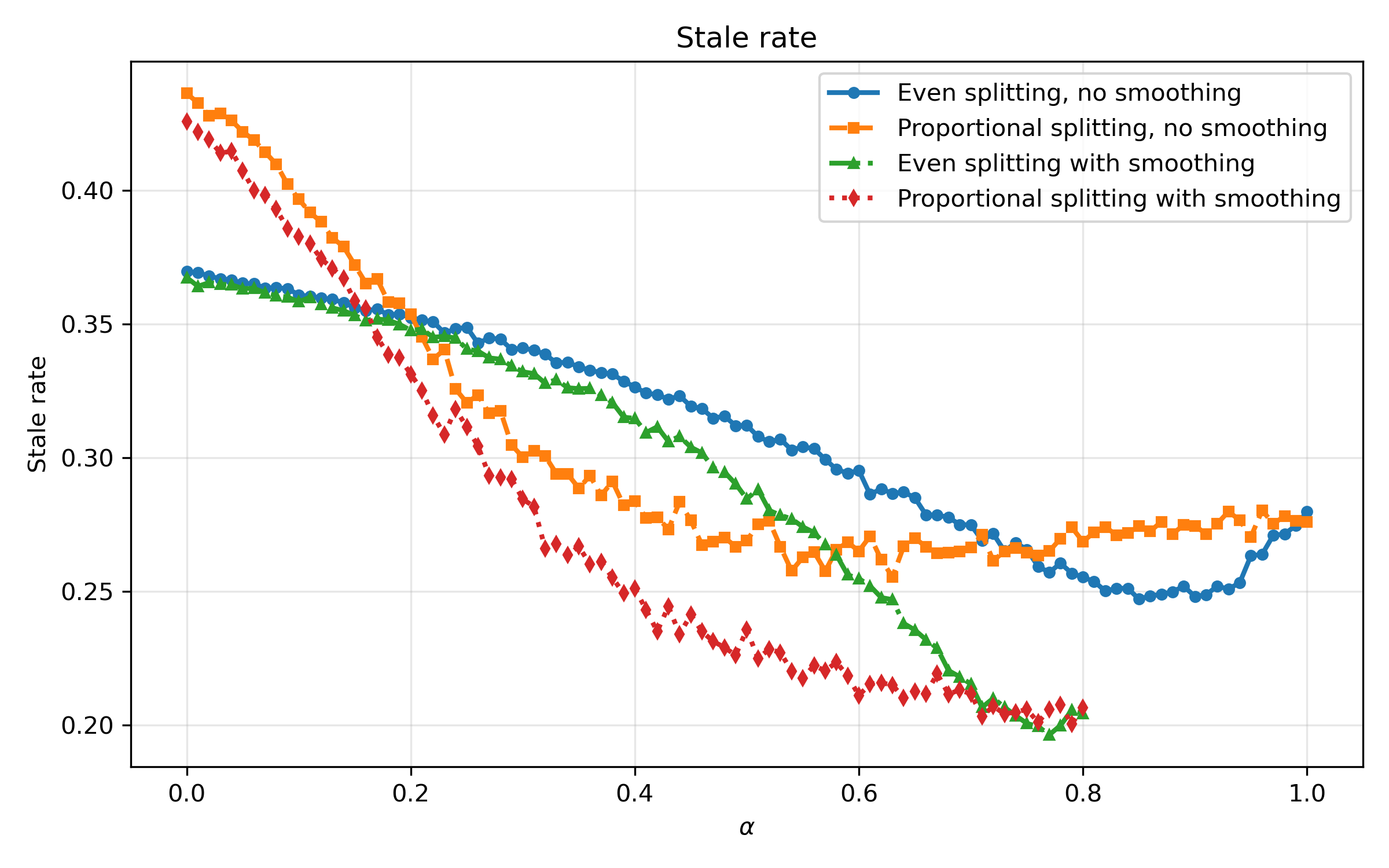}
    \caption{Stale block rate}
    \label{fig:throughput-b}
\end{subfigure}

\caption{System health metrics under fixed-strategy evaluation using the learned strategy distributions. Across the studied environments, canonical fee throughput remains close to one for a large range of $\alpha$. This indicates that the learned strategies generally do not prevent generated fee mass from reaching the canonical chain. The stale-rate curves show a more visible distinction: environments and/or parameter regions with greater fork-based behavior exhibit higher stale rates.}
\label{fig:throughput-panel}
\end{figure}

\paragraph{Results and interpretation}
The results, shown in Figure~\ref{fig:throughput-panel}, support the claim that not all fork-based learned strategies have the same system-level effect. When Function-Fork $1$ is dominant, as in Figure~\ref{fig:strategy-learning-b}, the system may still experience additional forks, but it does not appear to suffer substantial fee-inclusion loss. Thus, the harm from less aggressive forking strategies is better understood as wasted mining effort, measured by stale block rate, rather than as transaction-fee stagnation. More aggressive forking strategies remain more concerning as they can combine fork creation with intentional fee withholding. 

This interpretation is subject to the modeling abstraction inherited from Carlsten {\em et al.}~\cite{carlsten2016instability}: the simulator generates aggregate fee mass rather than individual transactions. In particular, it does not model transaction sizes, explicit transaction selection from the mempool, or block space constraints. The fee throughput metric should therefore be read as canonical inclusion of generated fee mass, not as literal transaction-count throughput or block space utilization.

\section{Conclusions, Limitations and Future Work}
In this work we presented a reward-splitting scheme with the goal of reducing incentives for fork-based mining strategies in PoW transaction-fee-only regimes. 
Analytically, we showed that deviation is profitable only when honest hash power is small relative to the winner-side reward 
and loser-side subsidy.  Experimentally, we identified parameter regions of this mechanism where Petty Honest behavior dominates fork-based deviations, and showed that even outside these regions network health remains higher than in the standard case.  
Our simulation-derived Default Compliance threshold for suppression of function-fork behavior significantly improves on both the $66\%$ bound identified by Carlsten {\em et al.} and the $30\%$ bound reported by Budinsk\'y {\em et al.}~\cite{budinsky2023fee} for FSRC. 
Since we also verified that orthodox selfish mining is not exacerbated by our mechanism, it appears to effectively mitigate undercutting incentives in the fee-only regime.

\textit{Limitations.} Our model uses a fixed $1\%$ hashpower threshold to determine atomicity. This is a useful modeling abstraction, but it cannot be directly implemented as stated: the true hashpower of a miner or mining pool cannot be observed a priori. Deployed systems would require an observable proxy for atomicity, e.g. recent block production. Designing such a proxy in a way that is robust to strategic splitting (of mining pools) or identity manipulation is an important direction for future work. 

Our strategy-learning experiments inherit the fixed strategy space studied by Carlsten et al. This allows for direct comparison with prior work, but does not rule out the existence of other deviant strategies. A richer deviant strategy space may reveal more powerful strategies, with corresponding effects on the level of Default-Compliance required to suppress them. Similarly, our empirical strategy-learning experiments use a fixed mining pool distribution. Future work could evaluate the sensitivity of our results to different distributions.

The reward mechanism also introduces several parameters: $\alpha, \beta, \delta, L, \gamma$. Although we explore these experimentally, we do not fully characterize the optimal parameter region or prove robustness across all possible environments. A more concrete analysis would identify parameter choices that simultaneously suppress undercutting, avoid encouraging selfish mining, and remain stable under dynamic mining power distributions. 

Finally, our selfish mining experiments consider only a single selfish miner competing against an honest baseline. Prior work shows that the dynamics of selfish mining can change when multiple selfish miners are present ~\cite{marmolejo2019competing, bai2023blockchain}. A natural direction for future work is therefore to study whether the reward mechanism continues to avoid worsening selfish mining incentives in the presence of multiple non-colluding selfish miners.

\bibliography{ref} 

\appendix


\section{Restricted-policy MILP formulation}
\label{app:milp}

This appendix gives the full restricted-policy MILP formulation discussed briefly in Section~\ref{subsec:restricted-milp-proxy}. The formulation is included as an analytical diagnostic and a possible extension of the closed-form threshold calculations. It is not used to generate the simulation results reported in Section~\ref{sec:simulation}. 

\subsection{MILP formulation over a restricted Function-Fork family}
\label{subsec:milp-equil}
We formulate a restricted-policy diagnostic for the atomic-player split model of Section~\ref{subsec:redistribution-chi}, specialized to the proportional loser-side subsidy rule. Fix an immediate-payment share \(\alpha\in(0,1)\), with \(\beta=1-\alpha\) in the no-smoothing case. For a fixed background mix \((p,\ell)\) of Petty Compliant and Lazy forkers, the MILP asks for the smallest Default Compliant mass under which the remaining strategic mass can be supported entirely on non-aggressive Function-Fork policies, with no profitable aggressive one-step deviation.

\paragraph{Restricted Function-Fork family.}
We discretize the Carlsten {\em et al.} \emph{Function-Fork} family by restricting the fee-claiming rule to
\[
\mathcal U=\Big\{1,\tfrac12,\tfrac14,\tfrac18\Big\},\qquad f_u(R)=uR,
\]
where $u$ is the fraction of currently available fee mass $R$ claimed by an undercutting miner; smaller $u$ corresponds to more aggressive undercutting. Let $\mathcal U_{\mathrm{aggr}}\subseteq\mathcal U$ denote the subset classified as aggressive; in the toy implementation below we use $\mathcal U_{\mathrm{aggr}}=\{\tfrac14,\tfrac18\}$. In the atomic-player split model, a fraction $\alpha$ of the realized fee revenue of the winning block is paid to the winner, while a fraction $\beta$ is redistributed only among \emph{non-winning} atomic miners that are eligible for the loser-side subsidy. Accordingly, the payoff of a strategic miner following policy $u$ has two components: a winner-side reward if its block is extended by the next miner, and a loser-side redistribution payment if its block is not extended and the miner remains an eligible non-winning atomic miner.

\paragraph{Scenario model and precomputation.}
Let $\Omega$ be a finite set of local blockchain states with weights $w_\omega\ge 0$ satisfying $\sum_{\omega\in\Omega} w_\omega=1$. For each $u\in\mathcal U$ and $\omega\in\Omega$, we precompute: (i) the included fee mass $\mathrm{inc}_u(\omega)$ if the block produced under policy $u$ becomes canonical; and (ii) the loser-side redistribution payment \(\rho_u(\omega)\geq 0\), computed using the same eligibility rule as in Section~4.3: only atomic miners are eligible for the subsidy, but the canonical winner may be atomic or non-atomic. Thus the proportional denominator is \(H_A-h_{W(\omega)}\mathbf 1\{W(\omega)\in A\}\), and the even-splitting denominator is \(|A|-\mathbf 1\{W(\omega)\in A\}\). (Here $\rho_u(\omega)$ is the scenario-level analogue of the loser-side quantity $d_i$ from Section~\ref{subsec:redistribution-chi}.) For each $\omega\in\Omega$, we also precompute one-step extension indicators $a_{\mathrm{Def},u}(\omega)$, $a_{\mathrm{Pet},u}(\omega)$, $a_{\mathrm{Lazy},u}(\omega)\in\{0,1\}$, together with $a_{v,u}(\omega)\in\{0,1\}$ for each $v\in\mathcal U$, where $a_{t,u}(\omega)=1$ iff a next miner of type $t$ extends the block produced by policy $u$ in scenario $\omega$. As in~\cite{carlsten2016instability}, this is a one-step local best-response model.

\paragraph{Variables and extension probabilities.}
Fix $(p,\ell)$ as parameters. Let $y$ denote the mass of \emph{DefaultCompliant} miners, and let $z_u\ge 0$ denote the mass of strategic miners using restricted policy $f_u$. The mass-balance constraint is
\[
y+p+\ell+\sum_{u\in\mathcal U} z_u = 1.
\]
Given $(y,p,\ell,z)$, the probability that a block produced by policy $u$ is extended in scenario $\omega$ is
\(
\pi_u(\omega;y,p,\ell,z)
= y\,a_{\mathrm{Def},u}(\omega) + p\,a_{\mathrm{Pet},u}(\omega) + \ell\,a_{\mathrm{Lazy},u}(\omega)
+ \sum_{v\in\mathcal U} z_v\,a_{v,u}(\omega)\).
Under the loser-side redistribution rule, if the block produced under policy $u$ is extended then the strategic miner receives $\alpha\,\mathrm{inc}_u(\omega)$; if it is not extended then it instead receives $\rho_u(\omega)$. Hence the expected payoff of policy $u\in\mathcal U$ is
\(
U_u(y,p,\ell,z)
= \sum_{\omega\in\Omega} w_\omega (
\alpha\,\mathrm{inc}_u(\omega)\,\pi_u(\omega;y,p,\ell,z)
+ \rho_u(\omega)(1-\pi_u(\omega;y,p,\ell,z))).
\)
Define $d_u:=\sum_{\omega\in\Omega} w_\omega\,\rho_u(\omega)$, and precompute
\[
c_u^{\mathrm{Def}} := \sum_{\omega\in\Omega} w_\omega\bigl(\alpha\,\mathrm{inc}_u(\omega)-\rho_u(\omega)\bigr) a_{\mathrm{Def},u}(\omega),\]
\[
c_u^{\mathrm{Pet}} := \sum_{\omega\in\Omega} w_\omega\bigl(\alpha\,\mathrm{inc}_u(\omega)-\rho_u(\omega)\bigr) a_{\mathrm{Pet},u}(\omega),
\]
\[
c_u^{\mathrm{Lazy}} := \sum_{\omega\in\Omega} w_\omega\bigl(\alpha\,\mathrm{inc}_u(\omega)-\rho_u(\omega)\bigr) a_{\mathrm{Lazy},u}(\omega),\]
\[
c_{u,v} := \sum_{\omega\in\Omega} w_\omega\bigl(\alpha\,\mathrm{inc}_u(\omega)-\rho_u(\omega)\bigr) a_{v,u}(\omega).
\]
Then the payoff-threshold equation admits the MILP-compatible linear form for \(u\in\mathcal U\)
\begin{equation}
\label{eq:payoff-linear-threshold-ymin}
U_u(y,p,\ell,z)
= d_u + y\,c_u^{\mathrm{Def}} + p\,c_u^{\mathrm{Pet}} + \ell\,c_u^{\mathrm{Lazy}} + \sum_{v\in\mathcal U} c_{u,v}z_v.
\end{equation}

\paragraph{No-aggressive restricted equilibrium constraints.}
Fix $(p,\ell)$. We seek the minimum $y$ such that the remaining strategic mass
can be allocated entirely to non-aggressive restricted policies and forms a
restricted equilibrium with no profitable aggressive deviation.
Thus we impose
\[
z_u=0 \qquad \forall u\in \mathcal U_{\mathrm{aggr}}.
\]
Let $U^*$ denote the restricted best-response payoff. We require
\[
U_u(y,p,\ell,z)\le U^* \qquad \forall u\in\mathcal U,
\]
and we enforce that every non-aggressive policy used in support attains this value. For $u\in\mathcal U\setminus\mathcal U_{\mathrm{aggr}}$, introduce binary variables $b_u\in\{0,1\}$ and a constant
\[
M:=\max_{u\in\mathcal U}\sum_{\omega\in\Omega} w_\omega
\max\{\alpha\,\mathrm{inc}_u(\omega),\rho_u(\omega)\},
\]
so that every feasible payoff lies in $[0,M]$. Then
\[
U^* \le U_u(y,p,\ell,z)+M(1-b_u)
\qquad \forall u\in\mathcal U\setminus\mathcal U_{\mathrm{aggr}}.
\]
Hence if $b_u=1$, then $U_u=U^*$. To link support to positive strategic mass, we use
\[
z_u\le b_u,\qquad z_u\ge \varepsilon b_u
\qquad \forall u\in\mathcal U\setminus\mathcal U_{\mathrm{aggr}},
\]
for a small constant $\varepsilon>0$. Therefore, any policy receiving positive strategic mass is non-aggressive and attains the restricted best-response payoff, while aggressive policies are not strictly better than the supported payoff level.

\paragraph{MILP.}
For each fixed \((p,\ell)\), the corresponding optimization problem is
\begin{align}
\min_{y,z,b,U^*}\quad & y \label{eq:milp-obj-ymin} \\
\text{s.t.}\quad
& y+p+\ell+\sum_{u\in\mathcal U} z_u =1, \label{eq:milp-mass-ymin} \\
& y\ge 0,\qquad z_u\ge 0 \quad \forall u\in\mathcal U, \label{eq:milp-nonneg-ymin} \\
& z_u = 0 \qquad \forall u\in\mathcal U_{\mathrm{aggr}}, \label{eq:milp-ban-aggr-ymin} \\
& U^* \ge U_u(y,p,\ell,z) \qquad \forall u\in\mathcal U, \label{eq:milp-upper-ymin} \\
& U^* \le U_u(y,p,\ell,z) + M(1-b_u) \quad \forall u\in\mathcal U\setminus\mathcal U_{\mathrm{aggr}}, \label{eq:milp-lower-ymin} \\
& z_u \le b_u \quad \forall u\in\mathcal U\setminus\mathcal U_{\mathrm{aggr}}, \label{eq:milp-support-upper-ymin} \\
& z_u \ge \varepsilon b_u \quad \forall u\in\mathcal U\setminus\mathcal U_{\mathrm{aggr}}, \label{eq:milp-support-lower-ymin} \\
& b_u\in\{0,1\} \quad \forall u\in\mathcal U\setminus\mathcal U_{\mathrm{aggr}}. \label{eq:milp-binary-ymin}
\end{align}
Constraints~\eqref{eq:milp-ban-aggr-ymin} and~\eqref{eq:milp-lower-ymin}--\eqref{eq:milp-binary-ymin} together ensure that any nonzero strategic
mass is placed only on non-aggressive policies that attain the restricted best-response payoff $U^*$, while
$U^*\ge U_u$ for all $u\in\mathcal U$ implies that aggressive policies are not strictly better than the supported non-aggressive ones (ties are permitted). Here $\varepsilon>0$ is chosen small relative to the grid resolution so that a selected support policy carries nonzero mass. The optimum value defines $y_{\min}(p,\ell;\alpha)$.

\begin{remark}[Adding smoothing]
If one adds a smoothing component \(\delta\) with \(\alpha+\beta+\delta=1\), the MILP structure is unchanged, but the precomputed payoff coefficients must be updated. The winner-side payoff for policy \(u\) becomes
\(
\alpha\,\mathrm{inc}_u(\omega)+s_u^W(\omega)
\), where \(s_u^W(\omega)\) is the released smoothing payment to the winner. The loser-side term \(\rho_u(\omega)\) must include both the current-fee subsidy and the smoothing-side loser subsidy, using the same atomic eligibility denominators as above. The resulting payoff remains linear after these scenario-level quantities are precomputed.
\end{remark}

\begin{remark}[Limitation]
This MILP threshold is not intended to be identical to the learning-based Default-Compliant threshold reported in Section~\ref{sec:simulation}. The MILP studies a one-step restricted-equilibrium condition over a discretized Function-Fork family, whereas the simulator reports the final learned strategy distribution under repeated no-regret updates. We therefore use the MILP as an analytical proxy for restricted local incentives, not as a direct prediction of the learned majority threshold.
\end{remark}

\section{Empirical Mining Distribution}
\begin{figure}[H]
\centering
\small
\setlength{\tabcolsep}{5pt}
\renewcommand{\arraystretch}{1.05}
\begin{tabular}{lrr lrr}
 \multicolumn{3}{c}{Atomic Miners}& \multicolumn{3}{c}{Non-Atomic Miners}\\
\hline
Mining entity & Blocks & Share &Entity & Blocks & Share\\ \hline
AntPool & 34,683 & 21.69\% & Unknown & 42,404 & 26.52\%\\
Foundry USA & 25,863 & 16.18\% & BTC.com & 1,583 & 0.99\%\\
ViaBTC & 18,923 & 11.84\% & Poolin & 1,145 & 0.72\%\\
F2Pool & 18,628 & 11.65\% & BTC M4 & 947 & 0.59\%\\
Binance Pool & 5,214 & 3.26\% & Ultimus & 676 & 0.42\%\\
Mara Pool & 4,600 & 2.88\% & Kucoin & 127 & 0.08\%\\
SBI Crypto & 2,468 & 1.54\% & Luxor & 62 & 0.04\%\\
Braiins Pool & 2,439 & 1.53\% & 1THash & 60 & 0.04\%\\
\end{tabular}
\caption{Empirical mining-pool distribution used in the strategy-learning experiments. Pool hash power is estimated from three years of historical block production data recorded at \cite{blockchain_pool_distribution}. Pools with estimated share at least \(1\%\) are designated atomic.  Five additional non-atomic pools contributed less than 0.03\% each. Blocks attributed to "Unknown" are split into identical non-atomic miners rather than treated as a single strategic mining entity.}
\label{fig:empirical-mining-distribution}
\end{figure}

\section{Strategy-learning convergence.}
\label{sec:conv-validation}
\begin{figure*}[t]
\centering
\begin{subfigure}[t]{0.24\textwidth}
    \centering
    \includegraphics[width=\textwidth]{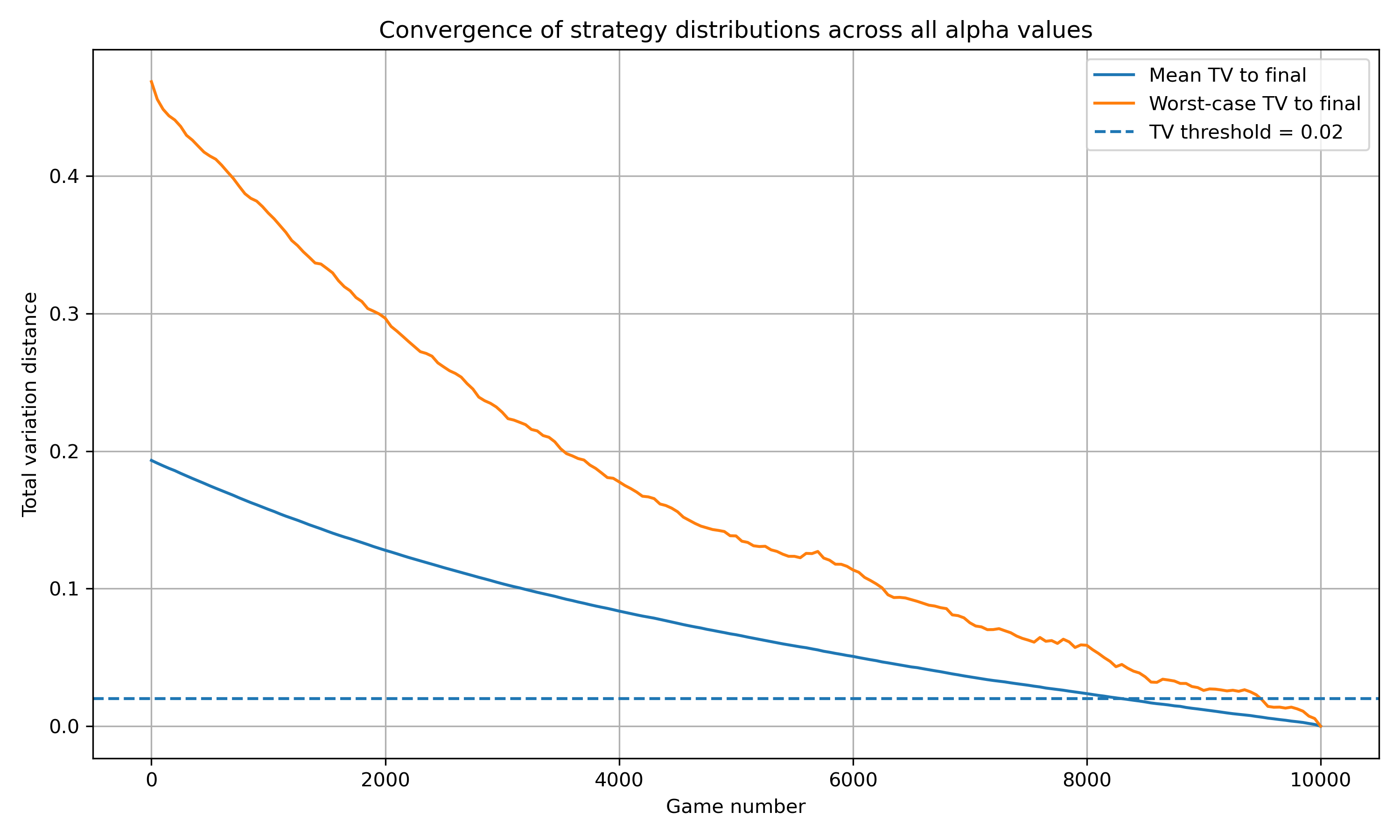}
    \caption{Even splitting, no smoothing.}
    \label{fig:even-no-smooth-conv}
\end{subfigure}
\hfill
\begin{subfigure}[t]{0.24\textwidth}
    \centering
    \includegraphics[width=\textwidth]{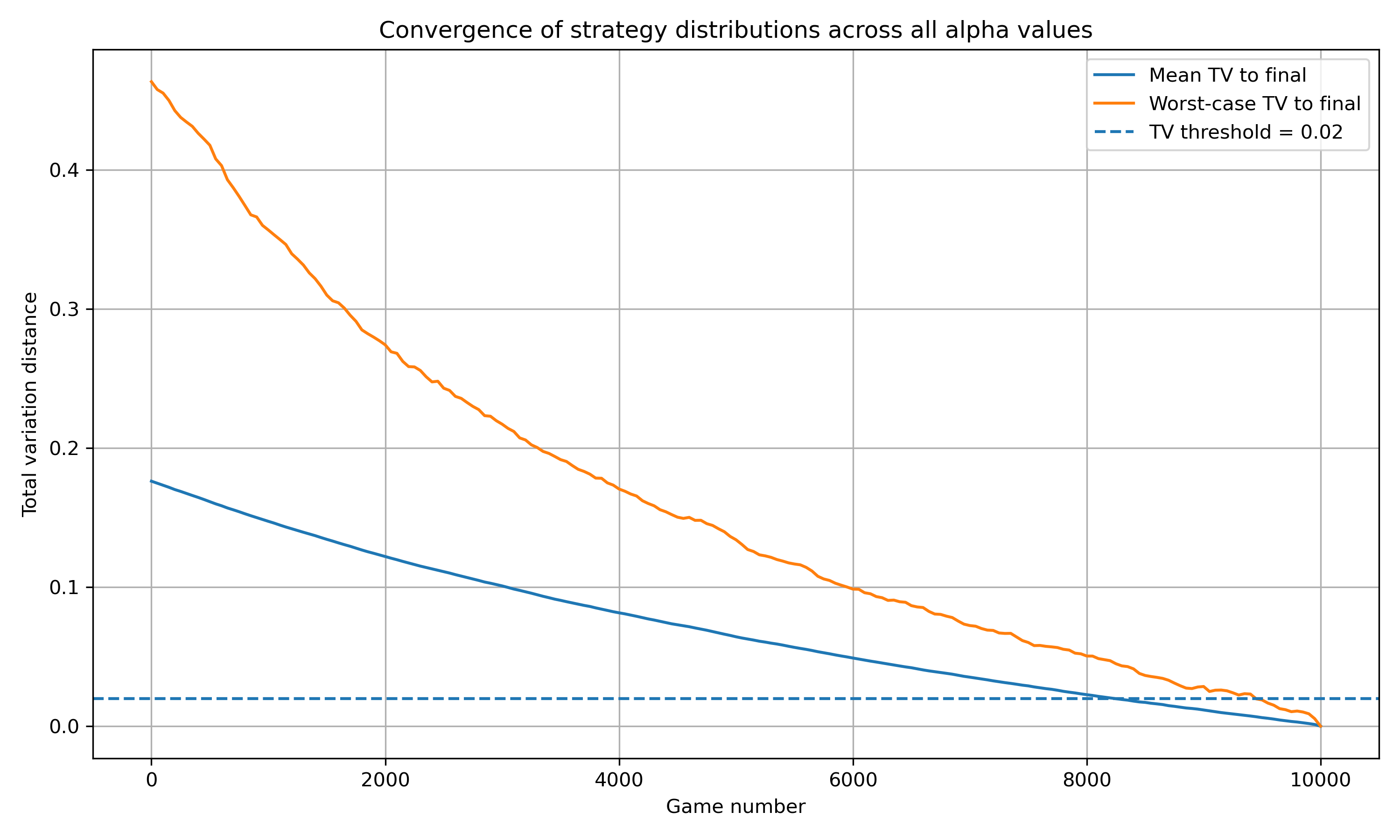}
    \caption{Even splitting with smoothing.}
    \label{fig:even-smooth-conv}
\end{subfigure}
\hfill
\begin{subfigure}[t]{0.24\textwidth}
    \centering
    \includegraphics[width=\textwidth]{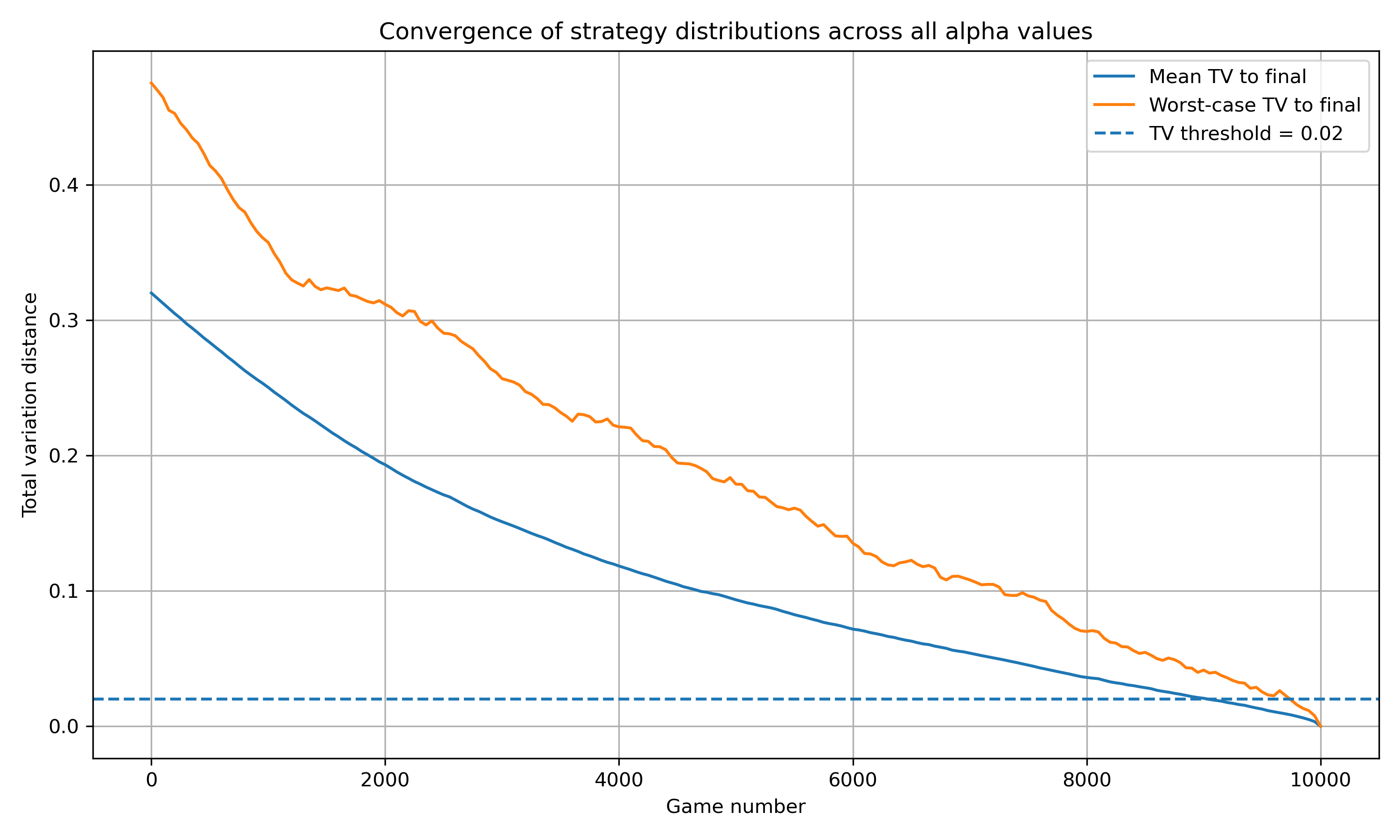}
    \caption{Proportional splitting, no smoothing.}
    \label{fig:prop-no-smooth-conv}
\end{subfigure}
\hfill
\begin{subfigure}[t]{0.24\textwidth}
    \centering
    \includegraphics[width=\textwidth]{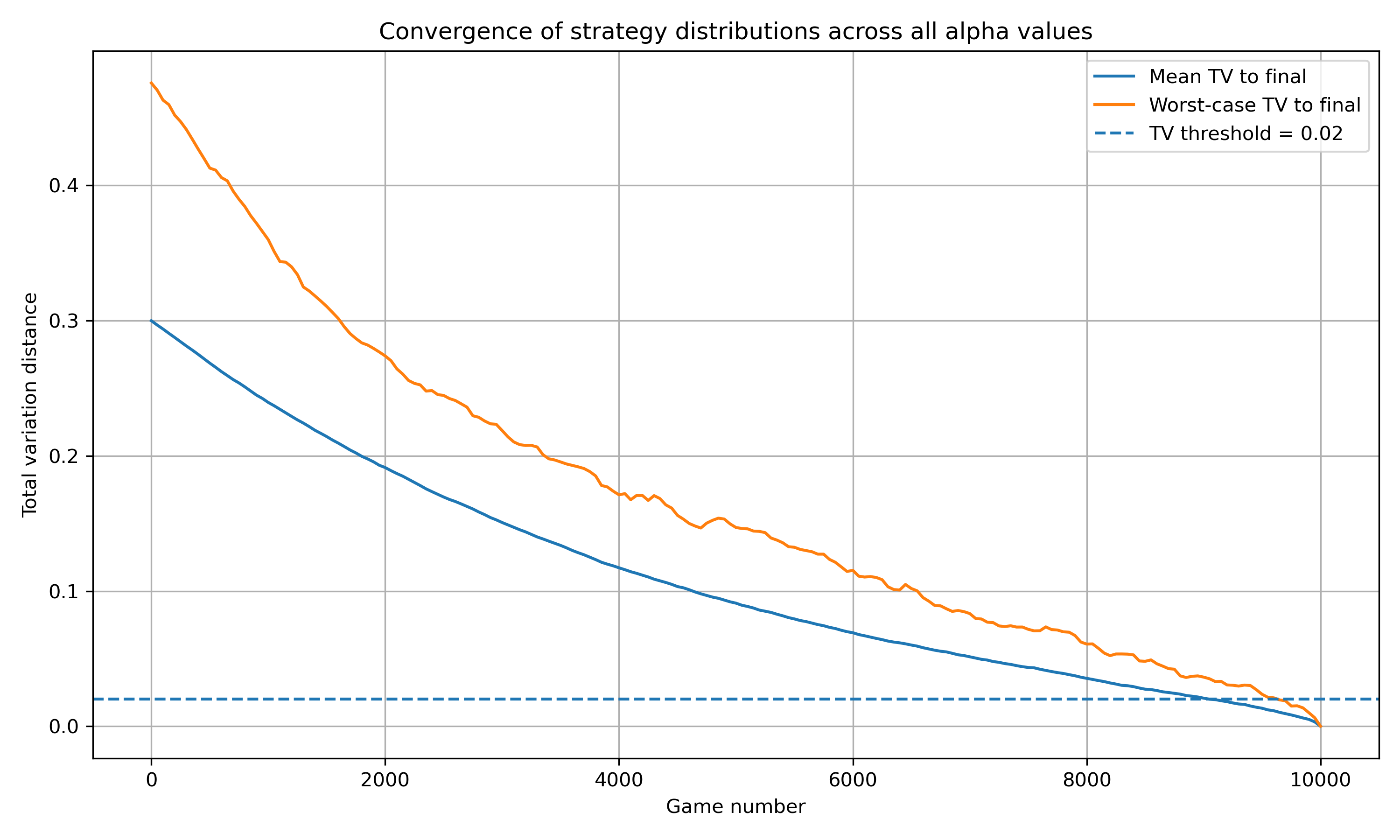}
    \caption{Proportional splitting with smoothing.}
    \label{fig:prop-smooth-conv}
\end{subfigure}

\caption{For each game number, our figures report both the average $\mathrm{TV}$ distance as well as the maximum $\mathrm{TV}$ distance over all analyzed values $\alpha$.}
\label{fig:convergence-panel}
\end{figure*}

The strategy learning experiments in section~\ref{subsubsec:strat-sim} report the learned strategy distribution after $10,000$ games, so we include a convergence check to verify that this endpoint is not simply an artifact of stopping the learning process early. For each fixed value of the reward parameter, $\alpha$, the simulator produces a strategy probability vector over the candidate strategy set, $\mathcal{S}$. We compare the vector at an intermediate game $t$, denoted $p_t^{\alpha}$, to the final vector at game $10,000$ for the same value of $\alpha$ denoted $p_{\mathrm{final}}^\alpha$. We produce total variation distance, or TV, plots; given by $\mathrm{TV}\left(p_t^\alpha, p_{\mathrm{final}}^\alpha\right) = \frac{1}{2}\left(\sum_{i \in \mathcal{S}} \left|p_{t,i}^\alpha - p_{\mathrm{final}, i}^\alpha\right|\right)$. The results can be seen in Figure~\ref{fig:convergence-panel}.

Across all convergence checks the learned strategy distributions stabilize before the $10,000$-game endpoint. Notably, both the average and maximum distances decrease towards zero with both falling below our $0.02$ threshold around $9,000$ games. This indicates that the reported final strategy distributions are not artifacts of an insufficient learning horizon, and supports using $10,000$ games for the strategy-learning experiments. 

Notably though, $10,000$ games were used for all environments: strategy-learning, default compliance, and selfish mining.

\section{A Concrete Bitcoin Instantiation}
\label{app:bitcoin-instantiation}

We describe one possible realization of the proposed rolling-window redistribution mechanism in Bitcoin. The purpose of this construction is to show that atomic miner designation and redistribution can be enforced using information already available in blocks and coinbase transactions, without requiring a separate smart contract system. 

Miners wishing to qualify for redistribution are required to use a consistent payment script for the first output of each coinbase transaction. We use the hash of this payment script as the miner identifier. Blocks produced using the same payment script can therefore be associated with the same miner identifier across the rolling window, without requiring the protocol to infer miner identities from network-level information. 

Once the block subsidy has fallen to zero, block validation is augmented with the following steps in addition to the existing consensus checks:

\begin{enumerate}
    \item The validating node computes the payments required under the selected splitting and smoothing rules. 
    \item The node examines the preceding $n$ accepted blocks and determines the set of atomic miners using the payment-script hashes appearing in their coinbase transactions. Under the rolling-window mechanism considered in our experiments, a miner is eligible if its identifier appears in at least the required number of blocks within this window. Under proportional redistribution, the number of appearances in the window additionally determines the miner's redistribution weight. 
    \item The node computes the payments required under the selected reward rule and verifies that the current block's coinbase transaction contains outputs paying the required amounts to the corresponding payment scripts. 
\end{enumerate}

A block that does not contain the required redistribution outputs is invalid under the modified consensus rules. Redistribution is therefore enforced as part of block validity rather than through an auxiliary contract or voluntary miner behavior. Since the relevant miner identifiers and rolling-window history are contained in the blockchain itself, every validating node can independently determine both the eligible miner set and the required payments. 

This construction has the advantage of requiring relatively little additional protocol machinery. Miner identifiers are represented by ordinary payment scripts, and eligibility is reconstructed directly from coinbase outputs. One limitation is that a miner's identity is tied to its payment script. If a miner changes this script, the protocol initially treats it as a new miner, causing the miner to temporarily lose the eligibility history accumulated under its previous identifier.

This construction directly implements the rolling-window splitting and smoothing redistribution mechanisms considered in our experiments.

We now provide a formal model:

Let $B[h]$ denote block $h$ in the blockchain. When verifying the validity of a new block $B[i]$, the relevant consensus checks can be summarized, in no particular order, as follows:
\begin{enumerate}
    \item \textit{Block-level cryptography}. Check that $B[i]$ references $B[i-1]$ as its previous block and that the hash of $B[i]$ is a valid puzzle solution
    
    \item \textit{Transaction-level cryptography}. Check that each transaction validates under the payment scripts of its inputs. 
    
    \item \textit{Transaction-level ledger checks}. Check that each transaction input has not been spent in a previous block $B[j]$, for $j \leq i$, and that the sum of the inputs is greater than or equal to the sum of the outputs.
    
    For the $t$-th transaction in candidate block $B[i]$, let
    \[F[i,t] \geq 0\]
    be the difference between its input value and output value, i.e. its transaction fee. Thus, the total transaction fees in block $B[i]$ would be
    \[F[i] = \sum_t F[i,t]\].
    
    \item \textit{Block-level ledger check}. Ensure that the sum of the outputs in the coinbase transaction does not exceed the block reward plus $F[i]$.
\end{enumerate}

To include splitting or smoothing, only step four requires modification. They are the following:

\begin{enumerate}
    \item[(a)] \textit{Atomicity}. For each block $B[i-n], \dots, B[i-1]$, collect the payment script of the first coinbase transaction output (TXO). Let $A$ be the set of scripts that appear at least $q$ times. Let $w_i$ denote the payment script of the first coinbase TXO in the candidate block $B[i]$, and let 
    \[A^{-} = A \setminus\{w_i\}\]
    be the set of eligible atomic miners other than the current block producer.

    \item[(b)] \textit{Smoothing only}. For each block in $B[i - L], \dots, B[i-1]$, compute the \textit{smoothing portion} of the block $i-k$ for $k=1, \dots, L$:
    \[S[i-k] = \delta F[i-k]\]
    Then compute block $i$'s \textit{smoothing payout} as:
    \[Z[i] = \frac{1}{L}\left( S[i-L] + \dots + S[i-1] \right)\]
    Let $P$ be the number of outputs in the coinbase transaction. Ensure that:
    \begin{enumerate}
        \item the coinbase transaction includes the smoothing payout $Z[i]$ to the producer of block $B[i]$; and
        \item the remaining outputs attributable to the current block's transaction fees sum to 
        \[(1-\delta)F[i]\]
        Equivalently, the total amount paid to the producer of $B[i]$ is 
        \[(1-\delta)F[i] + Z[i]\]
    \end{enumerate}

    \item[(c)] \textit{Splitting only}. Compute the redistribution subsidy of block $i$ as 
    \[ R[i] = \beta F[i]\] 
    Let $P \geq |A^-| + 1$ be the number of outputs in the coinbase TX. Ensure that:
    \begin{enumerate}
        \item the last $|A^-|$ TXOs are paid to the scripts in $A^-$, with each receiving 
        \[\frac{R[i]}{|A^-|}\]
        or, under proportional redistribution, the corresponding proportional share; and
        \item the first $P-|A^-|$ TXOs sum to 
        \[(1-\beta)F[i]\]
        Since this is the splitting only case, $\delta=0$, so $1-\beta=\alpha$.
    \end{enumerate}

    \item[(d)] \textit{Split'n'Smooth}. Compute $Z[i]$ and $R[i]$ as above.  Let $P \geq |A^-|+1$ be the number of outputs in the coinbase transaction. Ensure that:
    \begin{enumerate}
        \item the last $|A^-|$ TXOs are paid to the scripts in $A^-$ and for the amounts 
        \[\frac{R[i]}{|A^-|}\]
        or, under proportional redistribution, the corresponding proportional share; and
        \item the first $P-|A^-|$ TXOs include the smoothing payout $Z[i]$ to the producer of block $B[i]$ and sum to 
        \[(1-(\beta+\delta))F[i]+Z[i]\]
        Since $\alpha+\beta+\delta=1$, the second quantity is equivalently 
        \[\alpha F[i] + Z[i]\]
    \end{enumerate}
\end{enumerate}

\end{document}